\documentclass[11pt,twoside]{article}  
\usepackage[ansinew]{inputenc}
\usepackage{amsfonts}
\usepackage{amssymb,amsmath}
\usepackage{latexsym}

\newcommand {\debeq}	{\begin{eqnarray*}}
\newcommand {\fineq}	{\end{eqnarray*}}

\newcommand	{\intgen}	
{\int_0^\infty}

\newtheorem	{thm}		{Theorem}[section]

\newtheorem	{dfn}		[thm]{Definition}

\newtheorem     {rem}           {Remark}
\newtheorem	{prop}	[thm]{Proposition}
\newtheorem	{cor}		[thm]{Corollary}

\begin{document}

\title{The reconstructed tree in the lineage-based model of protracted speciation}
\author{\textsc{By Amaury Lambert, Hélène Morlon, Rampal S. Etienne}
}
\date{\today}
\maketitle
\noindent\textsc{%
Amaury Lambert {\rm(corresponding author)}\\
UPMC Univ Paris 06\\
Laboratoire de Probabilités et Modèles Aléatoires CNRS UMR 7599\\
And\\
Collège de France\\
Center for Interdisciplinary Research in Biology CNRS UMR 7241\\
Paris, France}\\
\textsc{E-mail: }amaury.lambert@upmc.fr\\
\textsc{URL: }http://www.proba.jussieu.fr/pageperso/amaury/index.htm\\
\textsc{Tel: } +33 1 44 27 85 69\\
\textsc{Fax: } +33 1 44 27 72 23\\

\noindent\textsc{%
Hélène Morlon\\
Center for Applied Mathematics \\
Ecole Polytechnique CNRS UMR 7641\\
Route de Saclay\\
91128 Palaiseau Cedex, France\\}

\noindent\textsc{%
Rampal S. Etienne\\
Community and Conservation Ecology\\
Centre for Ecological and Evolutionary Studies\\
University of Groningen\\
Box 11103\\
9700 CC Groningen\\
The Netherlands}

\begin{abstract}
\noindent

A popular line of research in evolutionary biology is the use of time-calibrated phylogenies for the inference of diversification processes. This requires computing the likelihood of a given ultrametric tree as the reconstructed tree produced by a given model of diversification. Etienne \& Rosindell (2012) \nocite{etienne2012prolonging} proposed a lineage-based model of diversification, called protracted speciation, where species remain incipient during a random duration before turning good species, and showed that this can explain the slowdown in lineage accumulation observed in real phylogenies. However, they were unable to provide a general likelihood formula. Here, we present a likelihood formula for protracted speciation models, where rates at which species turn good or become extinct can depend both on their age and on time. Our only restrictive assumption is that speciation rate does not depend on species status.

Our likelihood formula utilizes a new technique, based on the contour of the phylogenetic tree and first developed in Lambert (2010)\nocite{lambert2010contour}. We consider the reconstructed trees spanned by all extant species, by all good extant species, or by all representative species, which are either good extant species or incipient species representative of some good extinct species. Specifically, we prove that each of these trees is a coalescent point process, that is, a planar, ultrametric tree where the coalescence times between two consecutive tips are independent, identically distributed random variables. We characterize the common distribution 
of these coalescence times
%
%
%
in some, biologically meaningful, special cases for which the likelihood reduces to an elegant analytical formula or becomes numerically tractable.
\end{abstract}  	
\bigskip

\noindent
\textit{Running head.} The reconstructed tree in the protracted speciation model.

\medskip

\medskip

\section{Introduction}

A central question in evolutionary biology is to infer the nature of processes which have shaped the contemporaneous patterns of biodiversity. A popular approach is to use time-calibrated phylogenies of extant species (starting with Nee et al. 1994\nocite{nee1994reconstructed}) which have been independently built, e.g.,  from interspecific gene sequence information. The aim is to choose, among a class of models of speciation and extinction, the ones that are the most likely to have generated a given phylogeny, by maximum likelihood or Bayesian methods. One of the key steps in this process is to evaluate the likelihood of a phylogeny under a given model of diversification, for example where species are viewed as particles that can reproduce (speciation) or die (extinction) independently at random times. For such so-called \emph{lineage-based models} of diversification, it is elementary to compute the likelihood of the whole species tree, but it is a more complicated task to compute the likelihood of the species tree spanned by extant species, also called the \emph{reconstructed tree}. Reconstructed trees are ultrametric trees, in the sense that all tips are at the same distance to the root. The probability distribution of the reconstructed tree is well-known for the linear birth--death process of diversification, where lineages are assumed to reproduce and die independently, at exponential rates possibly varying in time (see Nee et al. 1994\nocite{nee1994reconstructed}, following the seminal work of Kendall 1948\nocite{kendall1948generalized}). This distribution is also known in the case of binomial sampling when only a fraction of extant lineages is sampled, independently with a certain fixed probability \cite{morlon2011reconciling,stadler2011mammalian,hallinan2012generalized}. A specific feature of the reconstructed tree generated by a linear birth--death process with binomial sampling is that its topology is uniform over topologies with ranked node splitting times, and that node splitting times, or node depths, are independent and identically distributed (iid). Ultrametric trees satisfying this property are called \emph{coalescent point processes} \cite{popovic2004asymptotic,aldous2005critical}. It is proved in Lambert (2010)\nocite{lambert2010contour} and Lambert \& Stadler (2013) that this result is robust to the Markov assumption, that is, it holds even if species lifetimes are not exponentially distributed, or otherwise put, when extinction rates may depend on the species age.\\

Etienne \& Rosindell (2012)\nocite{etienne2012prolonging} have proposed a lineage-based model of diversification, called \emph{protracted speciation model}, where newborn species are so-called \emph{incipient species} and become so-called \emph{good species} after some exponentially distributed time. This model is a lineage-based version of the individual-based protracted speciation model of Rosindell et al. (2010), and can explain the slowdown in lineage accumulation observed in real phylogenies, a phenomenon that could indeed be due to the fact that populations experiencing recent speciation are not detected as actual species. Alternative explanations include the dependence of speciation or extinction rates upon the overall number of species \cite{rabosky2008density,etienne2012diversity,etienne2012conceptual}, ecological speciation \cite{mcpeek2008ecological}, and geographic speciation \cite{pigot2010shape}.

Here, we consider a generalization of this model, where the times spent in the incipient stage (or in several incipient stages) and in the good stage can be correlated and have inhomogeneous and general distributions, that is, when the rates at which species can change type or become extinct may depend on time or on their age (and type). The interpretation of protracted speciation is that newly founded populations (i.e., incipient species) cannot be discriminated from their mother population before enough time has elapsed to complete genetic differentiation and/or reproductive isolation. In this view, all extant incipient species descending, by a chain of incipient species, directly from the same good species, are considered as a cloud of satellite populations belonging to the same species. This cloud must only have one \emph{representative species} in the phylogenetic tree. 
If the ancestor good species $a$ of the cloud is extant, then $a$ is the natural representative species of the cloud.
Otherwise, we set up a natural rule to define which of the extant descending incipient species of $a$ is the representative species. Roughly speaking, the one representative species of an extinct good species $a$ is chosen as the last 
 incipient species among species descending from $a$ by a chain of incipient species.\\ 

We study the reconstructed tree spanned by all extant species, by representative species and by good extant species (by decreasing order of inclusion). We prove that if the speciation rate does not depend on species status, then all three reconstructed trees are given by a  coalescent point process, and we provide numerical methods to compute the common distribution of node depths in each case. We also provide a closed formula in the case of the reconstructed tree spanned by all extant species, as well as by good extant species, in the original setting of Etienne \& Rosindell (2012)\nocite{etienne2012prolonging} when rates are age-independent and do not vary with time.\\

Hereafter, we will make a difference between the terms \emph{phylogenetic tree} and \emph{reconstructed tree}. The phylogenetic tree at time $T$ is the tree with edge lengths obtained after throwing away all points at distance larger than $T$ from the root (the future of $T$). The reconstructed tree is the tree obtained from the phylogenetic tree after removing all lineages that are extinct by time $T$.

In the next section, we specify the model assumptions, extending the constant rate model of Etienne \& Rosindell (2012)\nocite{etienne2012prolonging} in two directions: the homogeneous model, where rates can be stage-dependent, and the Markov model, where rates can be time-dependent. We also define the so-called ultimogeniture order on the (finite) set of species of the phylogenetic tree, and use this order to define the rule for the choice of representative species. In Section 3, we propose a first numerical method to compute the likelihood of the reconstructed trees, which is a natural follow-up to Etienne \& Rosindell (2012)\nocite{etienne2012prolonging}, involving an infinite set of coupled ordinary differential equations. In Section 4, we give the rigorous definition of a coalescent point process, we introduce a total order on the phylogenetic tree embedded in continuous time, and use this to prove that the reconstructed tree of all extant species is a coalescent point process. In Section 5, we propose two methods, one for the homogeneous model, and one for the Markov model, to compute the coalescent distribution for the reconstructed tree of all extant species. Each of these two methods can be applied to the constant rate model, resulting in the same closed formula. In Section 6, we adapt these two methods to the cases of good species and of representative species, both of which are much more efficient and accurate than the one given in Section 3. In Section 7, we discuss two extensions to our model: including several stages of incipientness, and assuming that only a fraction of extant species is sampled.

\section{Model and preliminaries}

\subsection{The protracted speciation model}

Following Etienne \& Rosindell (2012)\nocite{etienne2012prolonging}, we model the dynamics of a phylogeny by a time-continuous, (possibly) time-inhomogeneous, (possibly) non-Markovian, two-type process, where a birth event is interpreted as the arrival of an incipient species and a death event is interpreted as an extinction. We will always assume that species behave independently, that is, that there is no diversity-dependence (branching property). We assume that each species gives birth in a Poissonian manner, that is, with an instantaneous speciation rate $b$ which is a constant or nonconstant function of time.\\

Species can be of type 1 or 2, where 1 is the `incipient' stage and 2 is the `good' stage. Note that the case of several stages will be studied in the last section. \emph{The speciation rate is assumed to remain constant regardless of stages}. At speciation time, say $s$, the new species starts out in state 1. It remains in state 1 for a random duration $U_s$, which is the duration of the incipient stage. At time $t=s+U_s$, it can become extinct or change type, that is, turn into a good species, with a probability that may depend on $s$ and $U_s$. If it succeeds to turn into a good species, it then survives another random duration $V_s$, which is the duration of the good stage, after which it becomes extinct. In what follows, the random variables $U_s$ and $V_s$ may be correlated. We will first study the case when their distribution does not depend on $s$, a case referred to as the \emph{homogeneous model}, because then the dynamics of the diversification process is time-homogeneous. We will then focus on the case when the distribution of stages is given by instantaneous hazard rates, in which case the diversification process counting the numbers of species of all types is Markovian. This  case, referred to as the \emph{Markov model}, divides into the \emph{inhomogeneous Markov case}, where rates can be time-variable, and the \emph{homogeneous Markov case}, which is the model studied most by Etienne \& Rosindell (2012)\nocite{etienne2012prolonging}. We will term this latter case the \emph{constant rate model}. Note that the constant rate model is the intersection between the homogeneous model (where rates are time-independent but may be age-dependent) and the Markov model (where rates are age-independent but may be time-dependent).  

In the Markov model, an incipient species can become extinct at rate $\mu_1$ and can turn into a good species at rate $\lambda_1$; a good species becomes extinct at rate $\mu_2$. Note that these rates may be nonconstant functions of time (inhomogeneous Markov case). Regardless of species type, the speciation rate is $b$ (an assumption written $\lambda_1=\lambda_3$ in  Etienne \& Rosindell (2012), where numbers indexing species status were swapped). We will always make this assumption, but we stress that we do not make further assumptions on  the other parameters (for example we can very well have $\mu_1 \not= \mu_2$). We stress that the constant rate model studied in Etienne \& Rosindell (2012) indeed fits into our general framework assuming that $U$ is exponentially distributed with parameter $\nu_1:=\lambda_1+\mu_1$, that $V$ is independent of $U$, that $V$ equals 0 with probability $\mu_1/(\lambda_1+\mu_1)$, and otherwise follows the exponential distribution with parameter $\mu_2$. Note that then $E(U) = \nu_1^{-1}$ and $E(V) = (\lambda_1/\nu_1)\mu_2^{-1}$, so that the diversification process is supercritical (exponentially growing number of species with positive probability) iff $bE(U+V)>1$, that is, $b(\mu_2+\lambda_2) -\nu_1\mu_2>0$.
\\

In the general case, the event $\{V=0\}$ is the event that the species becomes extinct before turning good. If $U$ is set to 0, the process counting the number of species is a one-type Crump--Mode--Jagers process, as studied in Lambert (2009, 2010)\nocite{lambert2009allelic} and Lambert \& Stadler (2013)\nocite{lambertcoalescent}. In particular, if $U$ is set to $0$ and $V$ is exponentially distributed, then the process is a classical linear birth--death process, as studied in Nee et al. (1994), which is (homogeneous and) Markovian. It is known that in the aforementioned simple cases, the likelihood of the reconstructed tree can be put in product form, meaning that the coalescence times, or node depths, are independent. Actually, they are also equally distributed, so the reconstructed tree is a so-called coalescent point process (see below). We will show that the reconstructed tree (spanned by all extant species, or by all good extant species, or by all representative species) is again a coalescent point process, even if $U$ and $V$ are both truly random and possibly correlated, and even if their joint distribution is time-dependent.

\subsection{The ultimogeniture order and the definition of representative species}

From now on, we consider a protracted diversification process starting with one (incipient) progenitor species at time 0 and conditioned to have extant species at time $T$. We wish to endow its set of species, both extant and extinct, with a total order, regardless of types.\\

Recall that in our setting, at each speciation event, we discriminate between the mother species and the daughter species (a distinction that can be randomly defined in the Markov model, with equal probabilities for each of the two configurations). We can now define an order on the set of species, called \emph{ultimogeniture order}. In what follows, we will say that species $a$ is younger than species $b$ if $a$ was born later than $b$, in forward time.
\begin{dfn}
\label{dfn:order}
We define the \emph{ultimogeniture order}, denoted by the order relation $\prec$, as follows.
Let $a$ and $b$ be  two species  with most recent common ancestor species $c$. If $a=c$, we set $a\prec b$, and if $b=c$, we set $b\prec a$. Otherwise, we let $a'$ and $b'$ be the daughters of $c$ which are ancestors of $a$ and $b$ respectively. Then $a\prec b$ if $a'$ is younger than $b'$, and $b\prec a$ if $b'$ is younger than $a'$. We will most of the times say that $a$ is \emph{smaller} than $b$ instead of writing $a\prec b$.
\end{dfn}

Another way of defining this order is to recursively label each species by a finite word of integers as follows. The progenitor species is labeled $\varnothing$. Then, if $u$ is the label of a species born before $T$, then the youngest daughter of $u$ born before $T$ is labeled $u1$, its second youngest daughter born before $T$ is labeled $u2$, and so on. Then the ultimogeniture order is the lexicographical order associated with this labeling. 

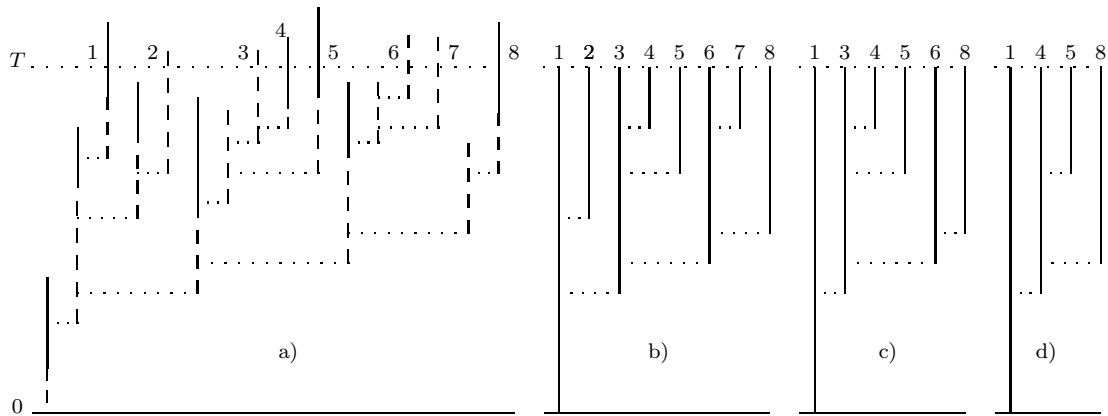
\begin{figure}[ht]
\unitlength 2mm 
\linethickness{0.4pt}
\ifx\plotpoint\undefined\newsavebox{\plotpoint}\fi 
\begin{picture}(74,30)(0,0)
\put(4,6){\line(0,1){6}}
\put(6,18){\line(0,1){4}}
\put(8,24){\line(0,1){5}}
\put(10,21){\line(0,1){4}}
\multiput(7.967,19.967)(-.6667,0){4}{{\rule{.4pt}{.4pt}}}
\multiput(11.967,18.967)(-.6667,0){4}{{\rule{.4pt}{.4pt}}}
\multiput(13.967,10.967)(-.88889,0){10}{{\rule{.4pt}{.4pt}}}
\multiput(5.967,8.967)(-.6667,0){4}{{\rule{.4pt}{.4pt}}}
\multiput(9.967,15.967)(-.8,0){6}{{\rule{.4pt}{.4pt}}}
\put(7.967,23.967){\line(0,-1){.8}}
\put(7.967,22.367){\line(0,-1){.8}}
\put(7.967,20.767){\line(0,-1){.8}}
\put(3.967,5.967){\line(0,-1){.75}}
\put(3.967,4.467){\line(0,-1){.75}}
\put(11.967,18.967){\line(0,1){.9}}
\put(11.967,20.767){\line(0,1){.9}}
\put(11.967,22.567){\line(0,1){.9}}
\put(11.967,24.367){\line(0,1){.9}}
\put(11.967,26.167){\line(0,1){.9}}
\multiput(13.967,16.967)(.6667,0){4}{{\rule{.4pt}{.4pt}}}
\put(15.967,16.967){\line(0,1){.875}}
\put(15.967,18.717){\line(0,1){.875}}
\put(15.967,20.467){\line(0,1){.875}}
\put(15.967,22.217){\line(0,1){.875}}
\multiput(15.967,20.967)(.6667,0){4}{{\rule{.4pt}{.4pt}}}
\put(17.967,20.967){\line(0,1){.875}}
\put(17.967,22.717){\line(0,1){.875}}
\put(17.967,24.467){\line(0,1){.875}}
\put(17.967,26.217){\line(0,1){.875}}
\put(21.967,18.967){\line(0,1){.8333}}
\put(21.967,20.633){\line(0,1){.8333}}
\put(21.967,22.3){\line(0,1){.8333}}
\put(22,24){\line(0,1){6}}
\put(23.967,12.967){\line(0,1){.875}}
\put(23.967,14.717){\line(0,1){.875}}
\put(23.967,16.467){\line(0,1){.875}}
\put(23.967,18.217){\line(0,1){.875}}
\put(24,20){\line(0,1){5}}
\multiput(23.967,20.967)(.6667,0){4}{{\rule{.4pt}{.4pt}}}
\put(25.967,20.967){\line(0,1){.8}}
\put(25.967,22.567){\line(0,1){.8}}
\put(25.967,24.167){\line(0,1){.8}}
\multiput(25.967,23.967)(.6667,0){4}{{\rule{.4pt}{.4pt}}}
\multiput(25.967,21.967)(.8,0){6}{{\rule{.4pt}{.4pt}}}
\put(27.967,23.967){\line(0,1){.8333}}
\put(27.967,25.633){\line(0,1){.8333}}
\put(27.967,27.3){\line(0,1){.8333}}
\put(29.967,21.967){\line(0,1){.8571}}
\put(29.967,23.681){\line(0,1){.8571}}
\put(29.967,25.395){\line(0,1){.8571}}
\put(29.967,27.109){\line(0,1){.8571}}
\multiput(23.967,14.967)(.88889,0){10}{{\rule{.4pt}{.4pt}}}
\put(31.967,14.967){\line(0,1){.8571}}
\put(31.967,16.681){\line(0,1){.8571}}
\put(31.967,18.395){\line(0,1){.8571}}
\put(31.967,20.109){\line(0,1){.8571}}
\multiput(31.967,18.967)(.6667,0){4}{{\rule{.4pt}{.4pt}}}
\put(33.967,18.967){\line(0,1){.8}}
\put(33.967,20.567){\line(0,1){.8}}
\put(33.967,22.167){\line(0,1){.8}}
\put(34,23){\line(0,1){6}}
\put(3,3){\line(1,0){32}}
\multiput(2.967,25.967)(.96875,0){33}{{\rule{.4pt}{.4pt}}}
\put(7,27){\makebox(0,0)[cc]{\scriptsize $1$}}
\put(11,27){\makebox(0,0)[cc]{\scriptsize $2$}}
\put(17,27){\makebox(0,0)[cc]{\scriptsize $3$}}
\put(23,27){\makebox(0,0)[cc]{\scriptsize $5$}}
\put(27,27){\makebox(0,0)[cc]{\scriptsize $6$}}
\put(31,27){\makebox(0,0)[cc]{\scriptsize $7$}}
\put(35,27){\makebox(0,0)[cc]{\scriptsize $8$}}
\put(38,3){\line(0,1){23}}
\put(55,3){\line(0,1){23}}
\put(68,3){\line(0,1){23}}
\put(40,26){\line(0,-1){10}}
\put(42,26){\line(0,-1){15}}
\put(57,26){\line(0,-1){15}}
\put(70,26){\line(0,-1){15}}
\put(46,26){\line(0,-1){7}}
\put(61,26){\line(0,-1){7}}
\put(48,26){\line(0,-1){13}}
\put(63,26){\line(0,-1){13}}
\put(74,26){\line(0,-1){13}}
\put(50,26){\line(0,-1){4}}
\put(52,26){\line(0,-1){11}}
\put(65,26){\line(0,-1){11}}
\multiput(36.967,25.967)(.9375,0){17}{{\rule{.4pt}{.4pt}}}
\multiput(39.967,15.967)(-.6667,0){4}{{\rule{.4pt}{.4pt}}}
\multiput(41.967,10.967)(-.8,0){6}{{\rule{.4pt}{.4pt}}}
\multiput(49.967,21.967)(-.6667,0){4}{{\rule{.4pt}{.4pt}}}
\multiput(51.967,14.967)(-.8,0){6}{{\rule{.4pt}{.4pt}}}
\put(37,3){\line(1,0){15}}
\put(38,27){\makebox(0,0)[cc]{\scriptsize $1$}}
\put(55,27){\makebox(0,0)[cc]{\scriptsize $1$}}
\put(68,27){\makebox(0,0)[cc]{\scriptsize $1$}}
\put(70,27){\makebox(0,0)[cc]{\scriptsize $4$}}
\put(40,27){\makebox(0,0)[cc]{\scriptsize $2$}}
\put(42,27){\makebox(0,0)[cc]{\scriptsize $3$}}
\put(57,27){\makebox(0,0)[cc]{\scriptsize $3$}}
\put(46,27){\makebox(0,0)[cc]{\scriptsize $5$}}
\put(61,27){\makebox(0,0)[cc]{\scriptsize $5$}}
\put(72,27){\makebox(0,0)[cc]{\scriptsize $5$}}
\put(48,27){\makebox(0,0)[cc]{\scriptsize $6$}}
\put(63,27){\makebox(0,0)[cc]{\scriptsize $6$}}
\put(50,27){\makebox(0,0)[cc]{\scriptsize $7$}}
\put(52,27){\makebox(0,0)[cc]{\scriptsize $8$}}
\put(65,27){\makebox(0,0)[cc]{\scriptsize $8$}}
\put(74,27){\makebox(0,0)[cc]{\scriptsize $8$}}
\multiput(56.967,10.967)(-.6667,0){4}{{\rule{.4pt}{.4pt}}}
\multiput(64.967,14.967)(-.6667,0){4}{{\rule{.4pt}{.4pt}}}
\multiput(53.967,25.967)(.91667,0){13}{{\rule{.4pt}{.4pt}}}
\put(54,3){\line(1,0){11}}
\multiput(66.967,25.967)(.875,0){9}{{\rule{.4pt}{.4pt}}}
\put(67,3){\line(1,0){7}}
\put(2,3.5){\makebox(0,0)[cc]{\scriptsize $0$}}
\put(2,26.5){\makebox(0,0)[cc]{\scriptsize $T$}}
\put(20,7){\makebox(0,0)[cc]{\scriptsize a)}}
\put(44.625,7){\makebox(0,0)[cc]{\scriptsize b)}}
\put(59.875,7){\makebox(0,0)[cc]{\scriptsize c)}}
\put(70.375,7){\makebox(0,0)[cc]{\scriptsize d)}}
\put(14,16){\line(0,1){8}}
\put(13.967,10.967){\line(0,1){.8333}}
\put(13.967,12.633){\line(0,1){.8333}}
\put(13.967,14.3){\line(0,1){.8333}}
\put(5.967,8.967){\line(0,1){.9}}
\put(5.967,10.767){\line(0,1){.9}}
\put(5.967,12.567){\line(0,1){.9}}
\put(5.967,14.367){\line(0,1){.9}}
\put(5.967,16.167){\line(0,1){.9}}
\put(9.967,15.967){\line(0,1){.8333}}
\put(9.967,17.633){\line(0,1){.8333}}
\put(9.967,19.3){\line(0,1){.8333}}
\multiput(21.967,18.967)(-.85714,0){8}{{\rule{.4pt}{.4pt}}}
\multiput(23.967,12.967)(-.90909,0){12}{{\rule{.4pt}{.4pt}}}
\multiput(17.967,21.967)(.6667,0){4}{{\rule{.4pt}{.4pt}}}
\put(19.967,21.967){\line(0,1){.6667}}
\put(19.967,23.3){\line(0,1){.6667}}
\put(20,24){\line(0,1){4}}
\put(19.5,28.5){\makebox(0,0)[cc]{\scriptsize $4$}}
\multiput(45.967,18.967)(-.8,0){6}{{\rule{.4pt}{.4pt}}}
\multiput(47.967,12.967)(-.85714,0){8}{{\rule{.4pt}{.4pt}}}
\multiput(41.967,21.967)(.6667,0){4}{{\rule{.4pt}{.4pt}}}
\put(44,22){\line(0,1){4}}
\multiput(60.967,18.967)(-.8,0){6}{{\rule{.4pt}{.4pt}}}
\multiput(62.967,12.967)(-.85714,0){8}{{\rule{.4pt}{.4pt}}}
\multiput(56.967,21.967)(.6667,0){4}{{\rule{.4pt}{.4pt}}}
\put(59,22){\line(0,1){4}}
\put(44,27){\makebox(0,0)[cc]{\scriptsize $4$}}
\put(59,27){\makebox(0,0)[cc]{\scriptsize $4$}}
\put(72,26){\line(0,-1){7}}
\multiput(71.967,18.967)(-.6667,0){4}{{\rule{.4pt}{.4pt}}}
\multiput(73.967,12.967)(-.8,0){6}{{\rule{.4pt}{.4pt}}}
\multiput(69.967,10.967)(-.6667,0){4}{{\rule{.4pt}{.4pt}}}
\put(40,27){\makebox(0,0)[cc]{\scriptsize$2$}}
\end{picture}
\caption{
a) A phylogenetic tree with two stages, starting from one ancestor species born incipient at time 0, with $8$ extant species at time $T$ labeled in the ultimogeniture order. Dotted lines indicate speciation events. Vertical edges start in dashed line, indicating the incipient stage, which sometimes turn into a solid line, indicating the good stage. The four species $2,3,6$ and $7$ are still incipient at time $T$, and species $2$ (resp. species $7$) is not representative, because the first extant descendant of its most recent good ancestor species is species $1$ (resp. species $6$); b) The reconstructed tree of all species extant at $T$; c) The reconstructed tree of representative species extant at $T$; d) The reconstructed tree of good species extant at $T$.
}
\label{fig : example}
\end{figure}

It is not difficult to see that this order defines a total order on any finite set of species, and in particular on the set of species, extant or extinct, of the tree stopped at time $T$. See Figure \ref{fig : example} for an example of a phylogenetic tree with 7 species extant at time $T$ labeled in the ultimogeniture order.

For any species $a$ with extant descendance, one can define the smallest extant descendant of $a$, or \emph{first extant descendant} of $a$ as the smallest species in the set of its extant descendant species, where `small' and `first' are to be understood in the sense of the ultimogeniture order. In other words, the first extant descendant of $a$ is the unique extant species $b$ descending from $a$ such that $b\prec c$ for any other extant species $c$ descending from $a$. Note that the first extant descendant of $a$ can also be defined recursively as the first extant descendant of its youngest daughter with extant descendance.
\\

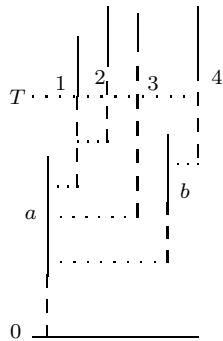
\begin{figure}[ht]
\unitlength 2mm 
\linethickness{0.4pt}
\ifx\plotpoint\undefined\newsavebox{\plotpoint}\fi 
\begin{picture}(35.25,25)(0,0)
\put(23,3){\line(1,0){11}}
\put(23.965,2.965){\line(0,1){.8333}}
\put(23.965,4.632){\line(0,1){.8333}}
\put(23.965,6.298){\line(0,1){.8333}}
\put(24,7){\line(0,1){8}}
\put(25.965,12.965){\line(0,1){.8571}}
\put(25.965,14.679){\line(0,1){.8571}}
\put(25.965,16.393){\line(0,1){.8571}}
\put(25.965,18.108){\line(0,1){.8571}}
\put(26,19){\line(0,1){4}}
\multiput(22.965,18.965)(.91667,0){13}{{\rule{.4pt}{.4pt}}}
\multiput(25.965,12.965)(-.6667,0){4}{{\rule{.4pt}{.4pt}}}
\multiput(25.965,15.965)(.6667,0){4}{{\rule{.4pt}{.4pt}}}
\multiput(23.965,10.965)(.85714,0){8}{{\rule{.4pt}{.4pt}}}
\put(29.965,10.965){\line(0,1){.9091}}
\put(29.965,12.783){\line(0,1){.9091}}
\put(29.965,14.601){\line(0,1){.9091}}
\put(29.965,16.419){\line(0,1){.9091}}
\put(29.965,18.238){\line(0,1){.9091}}
\put(29.965,20.056){\line(0,1){.9091}}
\put(27.965,15.965){\line(0,1){.8}}
\put(27.965,17.565){\line(0,1){.8}}
\put(27.965,19.165){\line(0,1){.8}}
\put(28,21){\line(0,1){4}}
\put(30,22){\line(0,1){2.5}}
\multiput(23.965,7.965)(.88889,0){10}{{\rule{.4pt}{.4pt}}}
\put(31.965,7.965){\line(0,1){.8}}
\put(31.965,9.565){\line(0,1){.8}}
\put(31.965,11.165){\line(0,1){.8}}
\put(33.965,14.965){\line(0,1){.8571}}
\put(33.965,16.679){\line(0,1){.8571}}
\put(33.965,18.393){\line(0,1){.8571}}
\put(33.965,20.108){\line(0,1){.8571}}
\put(32,12){\line(0,1){4.5}}
\put(34,21){\line(0,1){4}}
\multiput(33.965,14.465)(-.6667,0){4}{{\rule{.4pt}{.4pt}}}
\put(22,19){\makebox(0,0)[cc]{\scriptsize $T$}}
\put(21.875,3.375){\makebox(0,0)[cc]{\scriptsize $0$}}
\put(22.875,11.125){\makebox(0,0)[cc]{\scriptsize $a$}}
\put(33.125,12.75){\makebox(0,0)[cc]{\scriptsize $b$}}
\put(24.875,19.875){\makebox(0,0)[cc]{\scriptsize $1$}}
\put(27.5,20.25){\makebox(0,0)[cc]{\scriptsize $2$}}
\put(31,19.875){\makebox(0,0)[cc]{\scriptsize $3$}}
\put(35.25,20.25){\makebox(0,0)[cc]{\scriptsize $4$}}
\end{picture}
\caption{
 A phylogenetic tree with 4 extant species at time $T$. Species 1 is good at time $T$ but all other species are still incipient at $T$. This figure illustrates the fact that a species with extant descendants carrying the same allele can have no representative species. Species $a$ has no representative species at $T$ because its first extant descendant (species 1) is a good species, and so does not carry the same allele. However, species $2$ and $3$ are extant species both carrying the same allele as species 1. Species $b$ is represented by species $4$. 
}
\label{fig : counterexample}
\end{figure}

We now wish to define representative species. As in the infinite alleles model, assume that each new good species is given a new type, called \emph{allele} to avoid ambiguity with the stages, and assume this allele is inherited by all its daughter incipient species. As said in the introduction, all species with the same allele are seen as satellite populations of the same species that cannot be discriminated from each other, so that a phylogenetic tree cannot comprise more than one representative of all species sharing the same allele. We want to set up a rule to designate the representative species of each allele at time $T$. Then we will be able to consider the reconstructed tree of representative species, as the tree subtended by all species extant at $T$ that are representative of some good ancestor species. First, if the good ancestor species is extant at $T$, then it is naturally chosen as the representative species of its allele. If the good ancestor species is extinct by $T$, we should ideally designate the representative species as the smallest of the extant descendants of the good (extinct) ancestor species that share the same allele. With such a definition, any extinct good species with extant descending species sharing the same allele would be represented. However, it will be mathematically more convenient to set up the following alternative  rule, which is biologically less satisfying, because an extinct good species might have extant descending species sharing the same allele but no representative species (see Figure \ref{fig : counterexample} for an example).

\begin{dfn}
\label{dfn:represent}
Any good species extant at time $T$ is its own representative species. 
For any good species extinct at time $T$, if its first extant descendant shares the same allele, then it is designated as its representative species, otherwise no representative species is designated.
\end{dfn}
A consequence of this definition is that any extant incipient species which is the first extant descendant of some extinct good species is a representative species. See Figure \ref{fig : example}c for the reconstructed tree of representative species extant at $T$.

\section{An infinite set of coupled ODEs}

Let $P(n_1,n_2;t)$ denote the probability that one incipient species born at time $0$ has $n_1$ descendant incipient species and $n_2$ descendant good species at time $t$. Also let $P(\cdot,n_2;t)=\sum_{n_1\ge 0}P(n_1,n_2;t)$ denote the probability of $n_2$ descendant good species and $P(n_1,\cdot;t)=\sum_{n_2\ge 0}P(n_1,n_2;t)$ denote the probability of $n_1$ descendant incipient species.  

Etienne \& Rosindell (2012) provided an expression of the likelihood of a phylogeny with stem or crown age $T$ for the Markov version (with constant rates) of the protracted speciation model. This expression is a product of a multiplicative term involving $P(\cdot,0;T)$ and of evaluations of the function $f$ at node depths of the phylogeny (see next section), where $f(t)=P(\cdot, 1;T-t)$.

 The functions $P(\cdot,0;t)$ and $P(1,\cdot;t)$ can in principle be obtained by integrating the (infinite) system set of Kolmogorov differential equations satisfied by the functions $(P(n_1,n_2;t))_{n_1, n_2}$. When there is no extinction ($\mu _{1}=\mu
_{2}=0$), $P(n, 0 ;t)$ and $P(n, 1;t)$  can be computed analytically by solving the
corresponding partial differential equation for the probability generating
function (Etienne \& Rosindell 2012). 

When extinction is non-zero, however,
this trick no longer works because the partial differential equations are
not analytically tractable. A different trick, used by Kendall (1948) and Nee et al
(1994) is however possible, under the assumption $\mu _{1}=\mu _{2}=:\mu$. One
can then view the reconstructed birth-death process as a birth process with time-dependent
speciation initiation rate $ba_T(t)$ at time $T-t$ where $a_T(t)=(b-\mu)/\left(b-\mu e^{-\left( b-\mu\right)(T-t) }\right)$ is the probability of survival of the birth-death process with birth rate $b$ and death rate $\mu$, in $T-t$ time units. Thus, we get
\begin{subequations}
\begin{eqnarray}
\frac{d}{dt}P(n,0;t)  &=&ba_T(n-1) P(n-1,0;t) -( (
ba_T+\lambda _{1}) n) P(n,0;t)\\
\frac{d}{dt}P(n,1;t)  &=& ba_Tn
P(n-1,1;t) +\lambda _{1}( n
+1) P(n+1,0;t)\\ &-&( ba_T(
n+1) +\lambda _{1}n) P(n,1;t)\nonumber
\end{eqnarray}%
\end{subequations}
with initial conditions $P(n,0;0)=1$ if $n =1$ and $0$ otherwise, and $P(n,1;0)=0$ for all $n$. This procedure has three disadvantages. First, in
practice only an approximation can be used by truncating the infinite set of
ODEs at some arbitrary values of $n_1$ and $n_2$. Second,
the set of ODEs is large even for moderate upper limits of $n_1$ and $n_2$, and hence computationally demanding. Third, the
procedure is only valid under the assumption that $\mu _{1}=\mu _{2}$. In
this paper, we develop an approach which avoids these three disadvantages.

\section{Coalescent point processes and the reconstructed tree of all extant species}
\label{sec:cpp}

\subsection{Coalescent point processes}

\subsubsection{Definitions and main properties}

\begin{dfn}
A \emph{coalescent point process} is a random, planar, utrametric tree with edge lengths, where tips are numbered $0,1,2,\ldots$ from left to right, started with a single root point, and which satisfies the following two properties, monotonic labeling and independence, to be defined below. 
\end{dfn}
We call $T$ the stem age of this tree, that is, the common graph distance of tips to the unique root point. 

\begin{figure}[ht]
\unitlength 1.9mm 
\linethickness{0.2pt}
\begin{picture}(69,37)(-4,3)
\put(6,5){\vector(0,1){35}}
\put(4.961,5.961){\line(1,0){.9836}}
\put(6.928,5.961){\line(1,0){.9836}}
\put(8.895,5.961){\line(1,0){.9836}}
\put(10.863,5.961){\line(1,0){.9836}}
\put(12.83,5.961){\line(1,0){.9836}}
\put(14.797,5.961){\line(1,0){.9836}}
\put(16.764,5.961){\line(1,0){.9836}}
\put(18.731,5.961){\line(1,0){.9836}}
\put(20.699,5.961){\line(1,0){.9836}}
\put(22.666,5.961){\line(1,0){.9836}}
\put(24.633,5.961){\line(1,0){.9836}}
\put(26.6,5.961){\line(1,0){.9836}}
\put(28.568,5.961){\line(1,0){.9836}}
\put(30.535,5.961){\line(1,0){.9836}}
\put(32.502,5.961){\line(1,0){.9836}}
\put(34.469,5.961){\line(1,0){.9836}}
\put(36.436,5.961){\line(1,0){.9836}}
\put(38.404,5.961){\line(1,0){.9836}}
\put(40.371,5.961){\line(1,0){.9836}}
\put(42.338,5.961){\line(1,0){.9836}}
\put(44.305,5.961){\line(1,0){.9836}}
\put(46.272,5.961){\line(1,0){.9836}}
\put(48.24,5.961){\line(1,0){.9836}}
\put(50.207,5.961){\line(1,0){.9836}}
\put(52.174,5.961){\line(1,0){.9836}}
\put(54.141,5.961){\line(1,0){.9836}}
\put(56.108,5.961){\line(1,0){.9836}}
\put(58.076,5.961){\line(1,0){.9836}}
\put(60.043,5.961){\line(1,0){.9836}}
\put(62.01,5.961){\line(1,0){.9836}}
\put(63.977,5.961){\line(1,0){.9836}}
\multiput(4.961,32.961)(.983333,0){61}{{\rule{.4pt}{.4pt}}}
\put(4,33){\makebox(0,0)[cc]{\scriptsize $T$}}
\put(5.961,22.961){\line(1,0){1}}
\put(7.961,22.961){\line(1,0){1}}
\put(9.961,22.961){\line(1,0){1}}
\put(11.961,22.961){\line(1,0){1}}
\put(21.961,20.961){\line(1,0){1}}
\put(23.961,20.961){\line(1,0){1}}
\put(25.961,20.961){\line(1,0){1}}
\put(27.961,20.961){\line(1,0){1}}
\put(29.961,27.961){\line(1,0){1}}
\put(31.961,28.023){\line(1,0){1}}
\put(33.961,28.086){\line(1,0){1}}
\put(35.961,28.148){\line(1,0){1}}
\put(45.961,25.961){\line(1,0){1}}
\put(47.961,25.961){\line(1,0){1}}
\put(49.961,25.961){\line(1,0){1}}
\put(51.961,25.961){\line(1,0){1}}
\put(45.961,11.961){\line(-1,0){1}}
\put(43.961,11.961){\line(-1,0){1}}
\put(41.961,11.961){\line(-1,0){1}}
\put(39.961,11.961){\line(-1,0){1}}
\put(37.961,11.961){\line(-1,0){1}}
\put(35.961,11.961){\line(-1,0){1}}
\put(33.961,11.961){\line(-1,0){1}}
\put(31.961,11.961){\line(-1,0){1}}
\put(29.961,11.961){\line(-1,0){1}}
\put(27.961,11.961){\line(-1,0){1}}
\put(25.961,11.961){\line(-1,0){1}}
\put(23.961,11.961){\line(-1,0){1}}
\put(21.961,11.961){\line(-1,0){1}}
\put(19.961,11.961){\line(-1,0){1}}
\put(17.961,11.961){\line(-1,0){1}}
\put(15.961,11.961){\line(-1,0){1}}
\put(13.961,11.961){\line(-1,0){1}}
\put(11.961,11.961){\line(-1,0){1}}
\put(9.961,11.961){\line(-1,0){1}}
\put(7.961,11.961){\line(-1,0){1}}
\put(21.961,16.961){\line(-1,0){1}}
\put(19.961,16.961){\line(-1,0){1}}
\put(17.961,16.961){\line(-1,0){1}}
\put(15.961,16.961){\line(-1,0){1}}
\put(13.961,16.961){\line(-1,0){1}}
\put(11.961,16.961){\line(-1,0){1}}
\put(9.961,16.961){\line(-1,0){1}}
\put(7.961,16.961){\line(-1,0){1}}
\put(14,23){\line(0,1){10}}
\put(22,17){\line(0,1){16}}
\put(30,21){\line(0,1){12}}
\put(38,28){\line(0,1){5}}
\put(46,12){\line(0,1){21}}
\put(54,26){\line(0,1){7}}
\put(62,33){\line(0,-1){29}}
\put(14,34.5){\makebox(0,0)[cc]{\scriptsize 1}}
\put(22,34.5){\makebox(0,0)[cc]{\scriptsize 2}}
\put(30,34.5){\makebox(0,0)[cc]{\scriptsize 3}}
\put(38,34.5){\makebox(0,0)[cc]{\scriptsize 4}}
\put(46,34.5){\makebox(0,0)[cc]{\scriptsize 5}}
\put(54,34.5){\makebox(0,0)[cc]{\scriptsize 6}}
\thicklines
\put(9,33){\vector(0,-1){10}}
\put(17,33){\vector(0,-1){16}}
\put(25,33){\vector(0,-1){12}}
\put(33,33){\vector(0,-1){5}}
\put(41,33){\vector(0,-1){21}}
\put(49,33){\vector(0,-1){7}}
\put(11,28.75){\makebox(0,0)[cc]{\scriptsize $H_1$}}
\put(19,25.875){\makebox(0,0)[cc]{\scriptsize $H_2$}}
\put(27,28.125){\makebox(0,0)[cc]{\scriptsize $H_3$}}
\put(35,30.625){\makebox(0,0)[cc]{\scriptsize $H_4$}}
\put(43,20.625){\makebox(0,0)[cc]{\scriptsize $H_5$}}
\put(51,29){\makebox(0,0)[cc]{\scriptsize $H_6$}}
\end{picture}
\caption{Illustration of a coalescent point process showing the node depths $H_1,\ldots, H_6$ for each of the 6 consecutive pairs of tips. The node depth $H_7$ is the first one which is larger than $T$.
}
\label{fig : treecoal}
\end{figure}
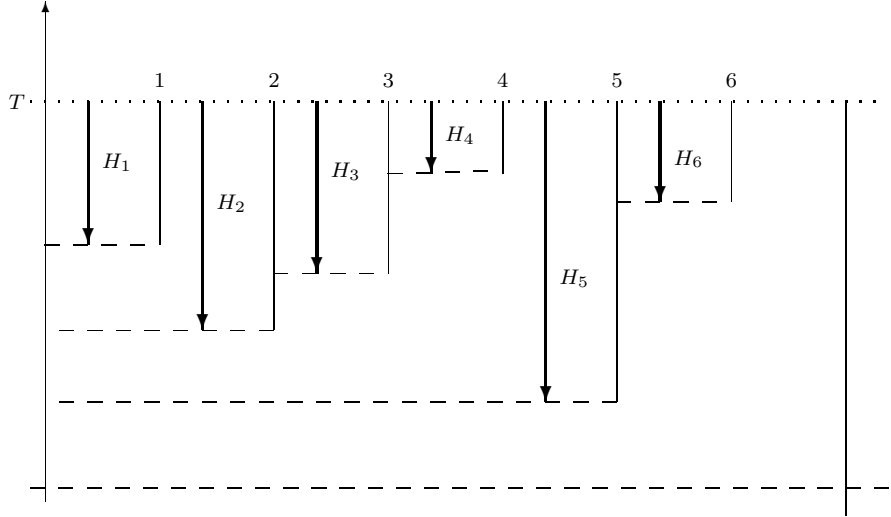

\paragraph{1. Monotonic labeling.} If $C_{i,i+k}$ denotes the \emph{coalescence time}, (or \emph{divergence time}) between tip $i$ and tip $i+k$, that is, the time elapsed since their lineages have diverged, then
\begin{equation}
\label{eqn : def coal}
C_{i,i+k}=\max\{H_{i+1},\ldots,H_{i+k}\}, 
\end{equation}
where $H_i:=C_{i-1,i}$.
In particular, the genealogical structure is entirely given by the knowledge of the sequence $H_1, H_2,\ldots$ that we will call either \emph{coalescence times} or \emph{node depths}. See Figure \ref{fig : treecoal} for a tree satisfying this property.

\paragraph{2. Independence.} There is a random variable $H$ (whose probability distribution may depend on $T$) such that node depths form a sequence of \textbf{independent, identically distributed random variables, all distributed as $H$,} killed at its first value larger than $T$.\\

Otherwise said, the number $N_T$ of tips in the coalescent point process follows the  geometric distribution with success parameter $P(H> T)$, and, conditional on $N_T=n$, the node depths $H_1,\ldots, H_n$ are independent copies of $H$ conditioned on $H\le T$. 
We will call the \emph{coalescent distribution} associated with a coalescent point process the law of $H$. It will often be convenient to use the inverse $W$ of the tail of the coalescent distribution as a way of characterizing it
$$
W(y) := \frac{1}{P(H>y)} \qquad y\ge 0.
$$
We will always assume that $H$ has a density (wrt Lebesgue measure), so that $W$ is differentiable and the density of $H$, say $f$, is given by
$$
f(y)=-\frac{d}{dy}P(H>y)=\frac{W'(y)}{W(y)^2}.
$$ 

\subsubsection{Likelihood formulae}

If a reconstructed tree has the law of a coalescent point process with coalescent density $f$ (given by $f=W'/W^2$, $W$ denoting the inverse of the tail of the coalescent distribution), then the likelihood ${\mathcal L}$ (conditional on at least 1 extant species) of a reconstructed tree $\tau$ with stem age $T$, $n$ extant species and node depths $x_1<\cdots< x_{n-1}$  is given by
\begin{equation}
\label{eqn:likelihood1}
{\mathcal L}(\tau)=\frac{C(\tau)}{W(T)}\prod_{i=1}^{n-1} f(x_i),
\end{equation}
where $C(\tau)$ is some combinatorial constant (see Tajima 1983). If ${\mathcal L}(\tau)$ is the mere likelihood of ranked node depths $x_1<\cdots< x_{n-1}$, then $C(\tau)= (n-1)!$, but if ${\mathcal L}(\tau)$ is the joint likelihood of the topology, or shape, of $\tau$, again with ranked node depths $x_1<\cdots< x_{n-1}$, then $C(\tau) = 2^{i(\tau)}$, where $i(\tau)$ is the number of nodes of $\tau$ that do not subtend cherries. 

Note that if $T$ is the \emph{crown age} of $\tau$, that is, if the two longest edges of $\tau$ both have length $T$, then the likelihood ${\mathcal L}_{\text{c}}(\tau)$ (the subscript `c' stands for `crown age') of the reconstructed tree $\tau$ with crown age $T$, $n$ extant species and node depths $x_1<\cdots< x_{n-2}$ (now there are only $n-2$ node depths strictly smaller than $T$), \emph{conditional on speciation at time 0 and survival of the two incident subtrees}, is  the product, properly renormalized, of the likelihoods of the two reconstructed subtrees conditional on survival, which equals 
\begin{equation}
\label{eqn:likelihood1bis}
{\mathcal L}_{\text{c}}(\tau)=\frac{C(\tau)}{W(T)^2}\prod_{i=1}^{n-2} f(x_i) ,
\end{equation}
where $C(\tau)$ was definde previously. This formula can be seen as obtained from the previous one by replacing one of the evaluations of $f$ by $W(T)^{-1}=P(H>T)$.  

Note that if the tree with stem (resp. crown) age $T$ is \emph{conditioned to have exactly $n$ tips}, then the conditioned likelihoods ${\mathcal L^n}$ (resp. ${\mathcal L^n_{\text{c}}}$) become
\begin{equation}
\label{eqn:likelihood2}
{\mathcal L^n}(\tau)= C(\tau) \prod_{i=1}^{n-1} f_T(x_i) \quad\mbox{ and resp. }\quad {\mathcal L^n_{\text{c}}}(\tau)= \frac{C(\tau)}{n-1} \prod_{i=1}^{n-2} f_T(x_i), 
\end{equation}
where $f_T(x)\, dx=P(H\in dx\mid H<T)$, that is, $f_T(x) = f(x) W(T)/(W(T)-1)$. Indeed, for the crown age, the probability to have $n$ tips conditional on two ancestors each having alive descendance at $T$ equals $(n-1)P(H<T)^{n-2} P(H>T)^2$.

Finally, we stress that all these likelihood formulae can be generalized to situations when not all extant species of the same clade are included in the tree. Indeed, most available phylogenies are not complete, in the sense that not all extant species descending from the same ancestor species are sampled and included in the phylogeny. In the next paragraph, we show how to compute the likelihood of the reconstructed tree of phylogenies which have missing extant species.

\subsubsection{Likelihood formulae with missing species}
\label{subsec:LF missing}

There are two main ways considered in the literature of randomly removing tips from a phylogenetic tree: the binomial model \cite{stadler2009incomplete,  lambert2009allelic,stadler2011mammalian,morlon2010inferring,morlon2011reconciling,hallinan2012generalized} and the $n$-sampling model \cite{stadler2009incomplete,etienne2012diversity}. Note that we can choose to first reconstruct the phylogenetic tree (i.e., throw away extinct lineages) and then remove tips from the reconstructed tree or first remove tips from the phylogenetic tree and then reconstruct the sampled tree, because both operations commute. In the $n$-sampling scheme, given a phylogenetic tree (or a reconstructed tree) with more than $n$ tips, $n$ tips are selected uniformly (e.g., sequentially) and all other tips are removed. In the binomial sampling scheme, or $\rho$-sampling scheme, given the phylogenetic tree (or the reconstructed tree), each tip is removed independently with probability $1-\rho$, where $\rho$ is the so-called \emph{sampling probability}. In Subsection \ref{subsec:missing}, we will also consider an extension of this sampling scheme where the sampling probability depends on the stage (incipient or good) of the tip species. \\


\paragraph{The $n$-sampling scheme.}

The tree obtained after $n$-sampling a coalescent point process is not a coalescent point process any longer. Assume we start from a coalescent point process with height $T$, with coalescent distribution given by some random variable $H$, and with a random number of tips  $N\ge n$. Select uniformly $n$ tips among $N$ (selecting uniformly one tip among $N$, then selecting uniformly a second tip among the remaining $N-1$, and so on $n$ times). Relabel the $n$ sampled tips $1, 2,\ldots, n$ ranked in the same order as they were in the initial coalescent point process and set $H_i'$ the coalescence time between sampled tip $i$ and sampled tip $i+1$, $i=1,\ldots, n-1$. By summing over all possible configurations of sampled tips, it is easy to see that for any $m\ge 0$ and any $x_1, \ldots, x_{n-1}\in [0,T]$
\begin{multline}
\label{eqn:big sum}
P(N=n+m, H_1' <x_1, \ldots, H_{n-1}'<x_{n-1}) = \frac{n! \ m!}{(n+m)!} \ P(H>T)\ \times\\ \times\ \sum_{\vec{m}:m_0+\cdots+m_n=m} P(H<x_1)^{m_1+1}\cdots P(H<x_{n-1})^{m_{n-1}+1} P(H<T)^{m_0+m_n},
\end{multline}
where the sum is taken over all possible vectors of non-negative integers $\vec{m}=(m_0,\ldots,m_n)$ such that $m_0+\cdots+m_n=m$. 

It is easy to differentiate \eqref{eqn:big sum} to get 
\begin{multline*}
P(N=n+m, H_1' \in dx_1, \ldots, H_{n-1}'\in dx_{n-1})/dx_1\cdots dx_{n-1}\\ = \frac{n! \ m!}{(n+m)!} P(H>T) \sum_{\vec{m}:m_0+\cdots+m_n=m}  P(H<T)^{m_0+m_n}\prod_{i=1}^{n-1}(m_i+1)f(x_i)P(H<x_i)^{m_i}  .
\end{multline*}
If we sum directly over all pairs $(m_0,m_n)$, and if we write $x_n= T$, we get
\begin{multline*}
P(N=n+m, H_1' \in dx_1, \ldots, H_{n-1}'\in dx_{n-1})/dx_1\cdots dx_{n-1}\\ = \frac{n! \ m!}{(n+m)!} P(H>T) \sum_{\vec{m}:m_1+\cdots+m_n=m}  (m_n+1)P(H<x_n)^{m_n}\prod_{i=1}^{n-1}(m_i+1)f(x_i)P(H<x_i)^{m_i}  .
\end{multline*}
In other words, the likelihood ${\mathcal L}^s(\tau)$ of a reconstructed tree $\tau$ with stem age $T$, $n$ \emph{sampled} species, $m$ missing species (i.e., $n+m$ extant species) and node depths $x_1<\cdots< x_{n-1}$, is given by (writing again $x_n=T$)
\begin{equation}
\label{eqn:likelihoodmissing}
{\mathcal L}^s(\tau)= {\mathcal L}(\tau) \ \frac{n! \ m!}{(n+m)!} \sum_{\vec{m}:m_1+\cdots+m_n=m}  \prod_{i=1}^{n}(m_i+1)P(H<x_i)^{m_i}
\end{equation}
where ${\mathcal L}(\tau)$ is given by \eqref{eqn:likelihood1}. A similar line of reasoning shows that the same correction factor holds for a reconstructed tree $\tau$ with \emph{crown age} $T$, $n$ sampled species, $m$ missing species and node depths $x_1<\cdots< x_{n-2}$, if now we write $x_{n-1}=x_n=T$.

\paragraph{The binomial sampling scheme.}
The $\rho$-sampling scheme  is trivial to handle in our situation. Indeed, as is explained in Lambert (2009) and Lambert \& Stadler (2013), the tree obtained after binomially sampling a coalescent point process with coalescent inverse tail distribution $W$ is, conditional on survival, a new coalescent point process with coalescent inverse tail distribution $W_\rho$ given by
$$
W_\rho = 1-\rho +\rho W.
$$ 
In Subsection \ref{subsec:missing}, we will extend these computations to cases when sampling probability depends on the stage of the species.

\subsection{Another total order}

 We wish to endow the phylogenetic tree associated with the diversification process with a total order which should be consistent with the total order on the set of species defined in Section 2. We stress that we think of the phylogenetic tree as a continuous object embedded in continuous time, whose elements are all timepoints belonging to edges of the tree, so that this order can be seen as a time-continuous process visiting all timepoints in the phylogenetic tree, which we call \emph{exploration process} (Lambert 2010). 
 
Recall that the phylogenetic tree at time $T$ is truncated at time $T$, in the sense that all the points at distance greater than $T$ from the root point are removed. The exploration process starts at the tip of the ancestor species' edge (at distance from the root equal to the extinction time of the ancestor species, or $T$, if the ancestor species is still extant at time $T$) and explores anterior points in this edge, running towards the root at unit speed, until it reaches the birth node of the youngest daughter species of the ancestor species born before $T$; at this time it jumps to the edge  tip of this daughter species (again, possibly truncated); when the exploration of an edge terminates (it always terminates at the birth node of this edge), it is immediately followed by the exploration of the mother edge at that node. The exploration is recursively defined in this way. This exploration process induces a total order on points of the phylogenetic tree, where the smallest element is the tip point of the ancestor edge and the largest element is its base point. In particular, the ultimogeniture order defined in Section 2 is also the order obtained when ranking species in the order where they appear in the exploration process.

\begin{figure}[ht]
\input{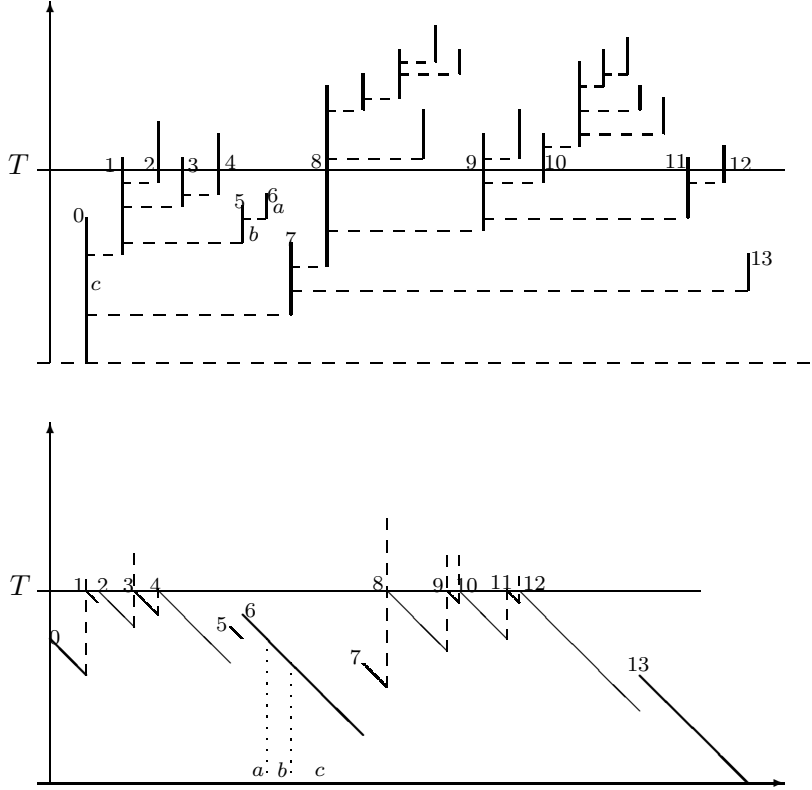}
\caption{ Top panel: A tree with edges in bold and speciation events shown by horizontal dashed lines (all horizontal edges have zero length); the species born before $T$ are labeled in the ultimogeniture order, and three zones of the tree are labeled by letters $a$, $b$ and $c$. Bottom panel: The contour process associated with the same tree after truncation at time $T$; edge labels are reported on top of each corresponding jump; epochs of visits of zones $a$, $b$ and $c$ by the contour process are indicated.}
\label{fig : jccp}
\end{figure}

The \emph{contour process} $X$ of the phylogenetic tree is a process living in $[0,T]$ and indexed by the same meaningless time variable as the exploration process. At any time $s$, $X_s$ is defined as the distance to the root of the point visited at time $s$ by the exploration process. Then the contour process has positive jumps (the lifetimes of species) and derivative $-1$ everywhere but at jump times (see Figure \ref{fig : jccp}). As can be seen in the figure, the contour process can be interpreted as the height of a ball that slips down the right-hand side of edges of the phylogenetic tree (embedded in the plane) at unit speed, and bounces back up to the next edge tip on its right each time it encounters a dashed line.\\

As was shown in Lambert (2010), this contour process $X$ is a Markov process which jumps at rate $b$ and makes jumps that are distributed as a species lifetime (that is, as $U+V$), which is truncated to $T$ when a jump overshoots $T$, and is killed when it hits 0. The number of visits of $T$ by this process is exactly the number of extant species at time $T$. Each time the contour process visits $T$, it makes a new excursion below $T$ which can either terminate by hitting 0 (end of the exploration) or by hitting $T$ (visit of a new extant species). Now recall that the excursions of a Markov process away from a given point (here, $T$) are independent and identically distributed (iid). Also observe by a quick inspection of Figure \ref{fig : jccp} that the coalescence time between two consecutive species visited by the contour process is exactly the depth of the excursion below $T$ starting at the first of these two visits and ending at the second one. This shows that, regardless of types, the reconstructed tree of all extant species at $T$ has iid node depths, all distributed as the depth of an excursion of $X$ below $T$. In particular, this ultrametric tree is a coalescent point process, whose coalescent distribution is the law of the depth of an excursion below $T$. 
We record this in the following statement.

\begin{prop}
\label{prop:all}
Under the protracted speciation model, conditional on at least one extant species at time $T$, the reconstructed tree spanned by all species extant at $T$ regardless of their types is a coalescent point process. The associated coalescent distribution is the law of the depth of an excursion away from $T$, made by the stochastic process $X$ which jumps at rate $b(s)$ when at $s$, with jump size distributed as $U_s+V_s$, and has slope $-1$ everywhere else. 
\end{prop}

Now with the last proposition in mind, equations \eqref{eqn:likelihood1} and \eqref{eqn:likelihood2} yield an expression for the likelihood of reconstructed trees of all extant species under the protracted speciation model, provided we can compute the associated coalescent distribution. The goal of the next section is to perform this computation. 


\section{Computation of the coalescent distribution for the tree spanned by all extant species}
\label{sec:all}

In this section, we treat in detail the case of the tree spanned by all extant species. We will build on these developments to give a more straightforward treatment of the trees, spanned by good species and by representative species, in the subsequent section.

From now on, we denote by $H$ the random variable associated with the coalescent point process of all extant species (see Proposition \ref{prop:all}), and we set
$$
W(y) := \frac{1}{P(H>y)} \qquad y\ge 0,
$$
the inverse of the tail of the coalescent distribution. We now show how to compute this function in the homogeneous model first, and then in the Markov model. We will then show how to use either of these methods to treat the constant rate model.

\subsection{The homogeneous model: a Laplace transform}

In the homogeneous model, neither $b$ nor the law of $(U,V)$ depend on time.
Then $W$ can be computed from the knowledge of the speciation rate $b$ and from the law of the total species lifetime $\Delta:=U+V$.

Let $\psi$ be the so-called Laplace exponent of the process $X$ in the homogeneous model.  The function $\psi$ is a convex function on $[0,\infty)$ that characterizes the law of $X$ and therefore only depends on the law of the total species lifetime $\Delta:=U+V$ and of the speciation rate $b$. More specifically,
$$
\psi(s) = s-b + b \,E(e^{-s\Delta})= s-b +b\, \intgen e^{-s x}\,P(\Delta \in dx)\qquad s\ge 0.
$$
Then it is known (Bertoin 1996) that 
$W$ is the unique non-negative function $g$ on $[0,\infty)$ satisfying
\begin{equation}
\label{eqn:Laplace}
\intgen e^{-sx} \, g(x) \, dx = \frac{1}{\psi(s)},
\end{equation}
for all $s$ greater than the exponential growth rate of the tree. The coalescent distribution can therefore be computed by inverting the previous Laplace transform. Indeed, recall from Section \ref{sec:cpp} that the density of the coalescent distribution, say $f$, is then given by $f(y)={W'(y)}/{W(y)^2}$.

\subsection{The Markov model: extinction probabilities}

Let us turn to the Markov model. Recall that in the Markov model, an incipient species can become extinct at rate $\mu_1$ and can turn into a good species at rate $\lambda_1$; a good species becomes extinct at rate $\mu_2$; the speciation rate is $b$, regardless of species type; all these rates may be nonconstant functions of time.

\paragraph{Convention.}
From now on, rates are expressed backwards from the stem age $T$, in the sense that $b(t)$ stands for the speciation rate at absolute time $T-t$, and similarly for other rates. In particular, $b(0)$ is the speciation rate at present time.\\
\\
The important idea behind the contour analysis is that for any integer $n$ smaller than the total number of tips: 1) all points of the tree visited after the visit of the ($n-1$)-th tip belong to subtrees that are independent of the past of the exploration process (the part of the tree visited before this visit), and 2) that those subtrees branch off the lineage joining the root to the ($n-1$)-th tip, in a Poissonian manner, with inhomogeneous intensity $b$, that is also independent from the past of the exploration process.


 Now the coalescence time between species $n-1$ and species $n$ is greater than $y$ if and only if all subtrees that have branched off this lineage at absolute times belonging in $(T-y, T)$, do not have descending species by time $T$. If $q_1(t)$ denotes the probability that a species in the incipient stage at absolute time $T-t$ has no extant descending species at absolute time $T$ (extinction probability), then $dt\,b(t)\, (1-q_1(t))$ is the probability that there is a subtree sprouting in the interval $(T-t, T-t+dt)$ and surviving up to  $T$, so that the zero-th term of the Poisson distribution of subtrees sprouting between $T-y$ and $T$ and surviving up to $T$ equals   
\begin{equation}
\label{eqn : all inhomogeneous case}
P(H > y) = \exp\left(-\int_0^y dt\,b(t)\, (1-q_1(t))\right)\qquad y\ge 0.
\end{equation}
 Then the problem moves to characterizing the function $q_1$. 
We do not have a closed formula for this function, but we know that the pair $(q_1, q_2)$ satisfies a system of Kolmogorov differential equations, where $q_2(t)$ is the probability that a  species in the good stage at absolute time $T-t$ has no extant descending species at absolute time $T$. This 2D differential equation is given by
\begin{equation}
\label{eqn : 2D extinction}
\left\{\begin{array}{rcl}
\dot{q}_1 &=& -(\nu_1+b) q_1 +\lambda_1 q_2 + \mu_1 + b q_1^2 \\
\dot{q}_2 &=& -(\mu_2+b) q_2 +\mu_2 + bq_1q_2,
\end{array}
\right.
\end{equation}
with initial conditions $q_1(0) = 0$ and $q_2(0)=0$. Recall that $\nu_1 = \lambda_1+\mu_1$, and that all rates $b$, $\lambda_1$, $\mu_1$, $\mu_2$ may depend on time.

Setting $g(y):=P(H>y)$ and recalling that $f$ is the density of $H$, we get $f=-\dot{g}$, and by \eqref{eqn : all inhomogeneous case}, 
\begin{equation}
\label{eqn:simple ODE}
\dot{g} = -b(1-q_1) g,
\end{equation}
so that 
\begin{equation}
\label{eqn:f}
f(y) = b(y)\, (1-q_1(y))\,\exp\left(-\int_0^y dt\,b(t)\, (1-q_1(t))\right)\qquad y\ge 0.
\end{equation}
If one numerically solves \eqref{eqn : 2D extinction}, then one can plug $q_1$ into \eqref{eqn:f} to get $f$ and hence the likelihood of any reconstructed tree, which is proportional to the product of evaluations of $f$ at node depths (see e.g. equation \eqref{eqn:likelihood1}). 
 To get $f$, one can equivalently integrate \eqref{eqn:simple ODE} (initial condition $g(0) =1$) simultaneously with \eqref{eqn : 2D extinction}, and then use $f=b(1-q_1)g$ to avoid computing the integral in \eqref{eqn:f}.
 
 If the dependence on $t$ by the instantaneous rates is piecewise constant as in Stadler (2011), then this numerical method should be particularly stable.
  
\begin{rem} 
Note that the second equation in \eqref{eqn : 2D extinction} can be integrated as 

$$
q_2(t) = \int_0^t \mu_2(y)\,dy\,\exp\left(-\int_y^t ds\,\mu_2(s)\right) \exp\left(-\int_0^{y} ds\,b(s)\, (1-q_1(s))\right).
$$
\end{rem}

\subsection{The constant rate model: computation of the coalescent distribution and of extinction probabilities}

Recall that the constant rate model can either be seen as a particular case of the homogeneous model, by specifying the probability distribution of $(U,V)$ as the one described in Section 2 ($U$ is exponentially distributed with parameter $\nu_1 := \lambda_1+\mu_1$, $V$ is independent of $U$, and either $V$ equals 0, with probability $\mu_1/(\lambda_1+\mu_1)$, or it is exponentially distributed with parameter $\mu_2$), or as a particular case of the Markov model, by assuming that rates are constant through time. 

Seeing the constant rate model as a particular homogeneous model, we can invert the Laplace transform in \eqref{eqn:Laplace} after specifying the distribution of $(U,V)$. Alternatively, seeing this model as a particular Markov model, we can compute the solution to \eqref{eqn : 2D extinction} with time-constant rates and plug the solution into \eqref{eqn : all inhomogeneous case}.\\

Let $Q$ be the following polynomial of degree 2
$$
Q(s) = s^2 + (\mu_2+\nu_1-b) s +\nu_1\mu_2 - b\mu_2 - b\lambda_1.
$$
It is easy to see that $Q$ always has two distinct real roots $\alpha < \beta$ given by
$$
\alpha = \frac{1}{2}\,\left(b-\nu_1-\mu_2 -\sqrt{K} \right)\quad \mbox{ and }\quad\beta = \frac{1}{2}\,\left(b-\nu_1-\mu_2 +\sqrt{K} \right),
$$
where $K:= (b+\mu_2-\nu_1)^2+4b\lambda_1$. It is also easy to see that $\alpha$ is always negative, so that $\beta> 0$ if and only if $\mu_2(\nu_1 - b) -b\lambda_1=\alpha\beta< 0$, that is, in the supercritical case. Actually, it can be shown that in this case, $\beta$ is the Malthusian parameter of the process counting the overall number of species (incipient or good). In other words, conditional on nonextinction, the overall  number of species grows exponentially with exponent $\beta$. Indeed, forthcoming equation \eqref{eqn:psi constant rate} shows that $\beta$ is the (only) positive root of $\psi$, which is shown in Lambert (2010) to be the Malthusian parameter of the corresponding branching process.

We now give a closed formula for $W$, which is the inverse of the tail of the coalescent distribution for the reconstructed tree of all extant species. 

\begin{prop}
When $\mu_2(\nu_1 - b) -b\lambda_1\not=0$, we have $\beta \not=0$ and then
\begin{equation}
\label{eqn : all Markov case}
W(y) = a_0 + a_1e^{\alpha y} + a_2 e^{\beta y},
\end{equation}
where 
$$
a_0 = \frac{\mu_2\nu_1}{\alpha \beta}, \quad
a_1 = \frac{(\alpha+\nu_1)(\alpha+\mu_2)}{\alpha(\alpha - \beta)}=\frac{b(\alpha+\lambda_1+\mu_2)}{\alpha(\alpha - \beta)},\quad
a_2 = \frac{(\beta+\nu_1)(\beta+\mu_2)}{\beta(\beta-\alpha)}=\frac{b(\beta+\lambda_1+\mu_2)}{\beta(\beta-\alpha)}.
$$
When $\mu_2(\nu_1 - b) -b\lambda_1=0$, we have $\beta=0$, $\alpha = b-\nu_1-\mu_2$, and
\begin{equation}
\label{eqn : all Markov case 2}
W(y)= b_0 + b_1e^{\alpha y} + b_2y e^{\alpha y},
\end{equation}
where
$$
b_0 = \frac{\nu_1\mu_2}{\alpha^2}, \quad
b_1 = 1-b_0,\quad
b_2 =\frac{b(b-\mu_1)}{\alpha}.
$$
\end{prop}
 We show this proposition by two different methods. Let us first use the results of the homogeneous model and proceed by Laplace transform inversion. With our distributions of $U$ and $V$, we get 
\begin{equation}
\label{eqn:Laplace lifetime}
E\left(e^{-s(U+V)}\right) = \frac{\mu_1(s+\mu_2)+\lambda_1 \mu_2}{(s+\nu_1)(s+\mu_2)},
\end{equation}
so that 
\begin{equation}
\label{eqn:psi constant rate}
\psi(s) = \frac{sQ(s)}{(s+\nu_1)(s+\mu_2)} .
\end{equation}
Elementary calculus then yields
$$
\intgen W(y)\, e^{-sy}\,dy=\frac{1}{\psi(s)} = \frac{(s+\nu_1)(s+\mu_2)}{sQ(s)}= \frac{a_0}{s}+\frac{a_1}{s-\alpha}+\frac{a_2}{s-\beta} .
$$
Depending on the signs of $a_0$, $a_1$ and $a_2$, we can invert this Laplace transform to get the announced result that $W(y) = a_0 + a_1e^{\alpha y} + a_2 e^{\beta y}$. The method is to substract all terms corresponding to negative coefficients among $a_0, a_1, a_2$, to equate the Laplace transforms of two positive functions and conclude by the injectivity argument. For example, if $a_2\le 0$ whereas $a_0, a_1\ge 0$, then $y\mapsto W(y)-a_2e^{\beta y}$ is a positive function whose Laplace transform equals $s\mapsto\frac{a_0}{s}+\frac{a_1}{s-\alpha}$, which is the Laplace transform of the positive function $y\mapsto a_0 + a_1 e^{\alpha y}$, hence the equality $W(y) - a_2 e^{\beta y}= a_0 + a_1e^{\alpha y} $.

We can do the same kind of calculations as previously in the case when $\beta=0$.\\

Let us now show how to apply the method developed for the Markov model. Note that here, because of time homogeneity, $q_1(t)$ (resp. $q_2(t)$) is the extinction probability in $t$ time units starting from one incipient species (resp. from one good species), regardless of the value of starting time. Recall from \eqref{eqn : all inhomogeneous case} that
$$
W(y) =\exp\left(b\,\int_0^y dt\, (1-q_1(t))\right),
$$
so that
\begin{equation}
\label{eqn:extinction incipient}
q_1 = 1- \frac{W'}{bW},
\end{equation}
an expression also displayed in Lemma 3.1 in Lambert (2011)\nocite{lambert2011species}. Plugging this into the first line in \eqref{eqn : 2D extinction}, we get
\begin{equation}
\label{eqn:extinction good}
q_2 = 1+ \frac{(b-\nu_1) W'-W''}{b\lambda_1 W}.
\end{equation}
Now if we plug the last two equalities into the second line of \eqref{eqn : 2D extinction}, we get
$$
W'''+(\mu_2+\nu_1-b) W''+(\nu_1\mu_2- b\mu_2 - b\lambda_1 ) W'= 0.
$$
Then $W'$ is the solution to a second-order, linear differential equation, whose characteristic polynomial is $Q$. As a consequence, $W'$ indeed is a linear combination of exponentials with exponents $\alpha$ and $\beta$, so that $W$ is a linear combination of exponentials with exponents $0$, $\alpha$ and $\beta$. We omit the detailed computation of the coefficients of this linear combination.\\

As a side result, we get the following expression for the extinction probabilities using \eqref{eqn:extinction incipient} and \eqref{eqn:extinction good}.
\begin{cor}
\label{cor:extinction}
In the constant rate model, as soon as $\mu_2(\nu_1 - b) -b\lambda_1\not=0$, the extinction probabilities $q_1(t)$ (resp. $q_2(t)$) in $t$ time units starting from one incipient species (resp. from one good species) are given by
$$
q_1(t) =  \frac{ba_0 + a_1 (b-\alpha) e^{\alpha t} + a_2 (b-\beta)e^{\beta t}}{b\left(a_0 +a_1e^{\alpha t} + a_2 e^{\beta t}\right)},
$$
and
$$
q_2(t) =  \frac{b\lambda_1a_0 + a_1\mu_2 (\alpha+\nu_1-b) e^{\alpha t} + a_2\mu_2 (\beta+\nu_1-b)e^{\beta t}}{b\lambda_1\left(a_0 +a_1e^{\alpha t} + a_2 e^{\beta t}\right)}.
$$
In the critical case, that is, when $\mu_2(\nu_1 - b) -b\lambda_1=0$, recall that $\beta =0$, $\alpha= b-\nu_1-\mu_2<0$, and we get
$$
q_1(t) =  \frac{bb_0 - (bb_0 +\alpha) e^{\alpha t}- b_2(\nu_1+\mu_2)te^{\alpha t}}{b\left(b_0 +b_1e^{\alpha t} + b_2 te^{\alpha t}\right)},
$$
and
$$
q_2(t) =  \frac{b\lambda_1b_0 - (b\lambda_1 b_0+\alpha b +\mu_2^2) e^{\alpha t} -b_2\mu_2^2 t e^{\alpha t}}{b\lambda_1\left(b_0 +b_1e^{\alpha t} + b_2 te^{\alpha t}\right)}.
$$
\end{cor}

\begin{rem} 

In the critical ($\beta=0$) and subcritical ($\beta<0$) cases, we see that the extinction probabilities increase exponentially fast to 1 as $t\to\infty$. In the supercritical case, $\beta$ is positive and the extinction probabilities converge as $t\to\infty$ to the overall extinction probabilities respectively equal to $1-(\beta/b)$ (when starting from one incipient species) and to $\mu_2(\beta+\nu_1-b)/(b\lambda_1)$ (when starting from one good species).
\end{rem}

\section{The reconstructed trees of good species and of representative species}

\subsection{The reconstructed tree of good species}

We show how to use the contour process to prove that the reconstructed tree of good species is a coalescent point process. We first make two observations. 

First, by the monotonicity property of the coalescent point process, the coalescence time between two good species $i$ and $j$ is the maximum of coalescence times of all consecutive pairs of species numbered $i,i+1,\ldots, j$. So if we can infer from the contour process which extant species are good and which extant species are incipient, we will be able to characterize the reconstructed tree of the good species. This the goal of our second observation, for which we need some notation. 

\begin{figure}[ht]
\unitlength 2mm 
\linethickness{0.4pt}

\begin{picture}(37,28)(-5,-7)
\thicklines
\multiput(22,8)(.01851852,-.01851852){54}{\line(0,-1){.01851852}}
\put(23,9){\line(1,-1){10}}
\put(33,5){\line(1,-1){2}}
\thinlines
\put(16,11){\line(1,-1){6}}
\put(13.5,11.5){\makebox(0,0)[cc]{$T$}}
\put(34.961,2.961){\line(0,1){.9375}}
\put(34.961,4.836){\line(0,1){.9375}}
\put(34.961,6.711){\line(0,1){.9375}}
\put(34.961,8.586){\line(0,1){.9375}}
\put(34.961,10.461){\line(0,1){.9375}}
\put(34.961,12.336){\line(0,1){.9375}}
\put(34.961,14.211){\line(0,1){.9375}}
\put(34.961,16.086){\line(0,1){.9375}}
\put(25,11){\vector(0,-1){12}}
\put(23.25,3){\makebox(0,0)[cc]{\scriptsize $H$}}
\put(35.5,9){\line(1,0){1}}
\put(36,9){\vector(0,-1){6.25}}
\put(37.75,5.625){\makebox(0,0)[cc]{\scriptsize $U$}}
\put(36,9){\vector(0,1){8}}
\put(37.75,12.75){\makebox(0,0)[cc]{\scriptsize $V$}}
\put(15,11){\line(1,0){30.75}}
\put(42,11){\vector(0,-1){8.25}}
\put(43.875,7.375){\makebox(0,0)[cc]{\scriptsize $A$}}
\put(16,-5){\vector(0,1){22}}
\put(15,-5){\vector(1,0){32}}
\multiput(34.961,2.961)(.90909,0){12}{{\rule{.4pt}{.4pt}}}
\multiput(24.961,-1.039)(.88889,0){10}{{\rule{.4pt}{.4pt}}}
\end{picture}
\caption{An excursion of the contour process away from $T$, showing its depth $H$, the undershoot $A$ of its terminating jump, which is the age of the corresponding extant species (the species whose lifespan traverses time $T$), the age $U$ at which it turned good, and the time $V$ is survived after turning good. In this example, this extant species is a good species at time $T$, because $U\le A$.}
\label{fig : excursion}
\end{figure}
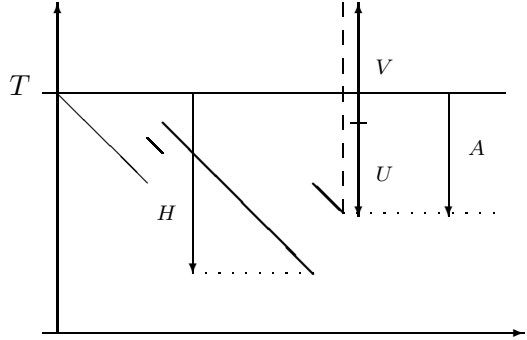

For each species extant at $T$, we call $A$ its age at time $T$ and $U$ its age when it turns good (see Figure \ref{fig : excursion}). If $U>A$, then the species is still incipient at $T$, otherwise it is a good extant species. In terms of the contour process, an extant species corresponds to a jump starting below $T$ and ending above $T$ (before truncation). The age $A$ of the corresponding extant species is called the `undershoot' of this jump. Because the triple of the depth $H$ of the excursion, of the age $A$ of the species by which the excursion ends and of its age at maturity $U$, is a function of the same excursion, all the triples $(H,A,U)$ running over all extant species, are independent and identically distributed.\\

Excursions satisfying $U\le A$ correspond to good species. Between two consecutive such excursions, we have a (geometric) number of excursions satisfying $U<A$, and we have to take the maximum of their depths $H$ to get the coalescence time between the two consecutive good species. If we denote by $H^g$ the associated random variable, we get 
$$
P(H^g <y) = \sum_{n=0}^\infty P(H<y, U>A)^n \,P(H<y, U\le A),
$$
because $H^g <y$ if and only if all the depths of in-between excursions (terminating with a species which is still incipient at $T$) are smaller than $y$. 
This can be recorded in the following statement.

\begin{prop}
\label{prop : good cpp}
Conditional on at least one good species extant at time $T$, the reconstructed tree spanned by extant good species is a coalescent point process. Its associated coalescent distribution is characterized by
\begin{equation}
\label{eqn : good cpp}
P(H^g <y) = \frac{P(H<y, U\le A)}{1- P(H<y, U> A)}.
\end{equation}
\end{prop}
Recall from \eqref{eqn:likelihood1} and \eqref{eqn:likelihood2} that the knowledge of the coalescent distribution is sufficient to compute the likelihood of the reconstructed tree of  extant good species under the protracted speciation model. We now show how to perform this computation, in the same vein as in the previous section.

\subsection{Computation of the coalescent distribution for the tree spanned by  good extant species}

The following statement shows how to recover the law of $H^g$ in the homogeneous model. Recall that the inverse $W^g$ of the tail of the coalescent distribution of good species is defined by $W^g(y) = 1/P(H^g>y)$ and that the inverse $W$ of the tail of the coalescent distribution associated with the tree spanned by all extant species can be recovered by inverting the Laplace transform  \eqref{eqn:Laplace}.
\begin{prop}
\label{prop : good cpp 2}
In the homogeneous model, the function $W^g$ is given by 
\begin{equation}
\label{eqn : good cpp 2}
W^g(y) = W(y) - b\,\int_0^y W(y-x)\,P(U>x)\, dx .
\end{equation}
\end{prop}
Note that the obvious inequality $W_g\le W$ implies that $P(H^g >y) \ge P(H>y)$ for all $y$, confirming that the node depths of the good species tree are larger than the node depths of the reconstructed tree of all extant species. 
\paragraph{Proof.}
 We use the following  distributional equation (see e.g. \cite{kyprianou2006introductory, lambert2011splitting}) characterizing the joint law of the depth $H$ of an excursion, of the size $D$ of the jump terminating this excursion, and of the undershoot $A$ of this jump, in terms of the law of the total species lifetime $\Delta= U+V$
$$
P(H<y, A\in dx,  D \in dz) = b \,\frac{W(y-x)}{W(y)}\, dx \,P(\Delta \in dz)  \qquad 0\le x \le \min(y,z).
$$
Then we get
\begin{eqnarray*}
P(H<y, U<A) &=& \int_{x=0}^y \int_{z=x}^{\infty} P(H<y, A\in dx, U<x,  D\in dz)\\
							&=& \int_{x=0}^y \int_{z=x}^{\infty} P(H<y, A\in dx, D\in dz)\,P(U<x\,|\, U+V =z)\\
							&=& b \,\int_{x=0}^y \int_{z=x}^{\infty} \frac{W(y-x)}{W(y)} \,P(\Delta \in dz)\,P(U<x\,|\, U+V =z)\, dx\\		
							&=& b \,\int_{x=0}^y \int_{z=x}^{\infty} \frac{W(y-x)}{W(y)}\,P(U<x, U+V \in dz)\, dx\\		
							&=& b \,\int_{x=0}^y \frac{W(y-x)}{W(y)}\, P(U<x, U+V >x)\, dx.	
	\end{eqnarray*}
Similarly, we obtain
$$
P(H<y, U>A ) = b\,\int_0^y \frac{W(y-x)}{W(y)}\,P(U>x)\, dx.
$$
Equation \eqref{eqn : good cpp 2} then  simply stems from plugging the last two equalities into \eqref{eqn : good cpp}.\hfill $\Box$\\

For the Markov model, we can repeat the same argument as that given in the previous section, to get 
\begin{equation}
\label{eqn : good inhomogeneous case}
P(H^g > y) = \exp\left(-\int_0^y dt\,b(t)\, (1-p_1^g(t))\right)\qquad y\ge 0,
\end{equation}
where $p_1^g(t)$ is the probability that a species in the  incipient  stage at absolute time $T-t$ has no \emph{good} descending species extant at absolute time $T$. Again, we do not have a closed formula for $p_1^g$, but the pair $(p_1^g, p_2^g)$ satisfies the following system of Kolmogorov differential equations, where $p_2^g(t)$ is the probability that a species in the good stage at absolute time $T-t$ has no good descending species extant at absolute time $T$. 
\begin{equation}
\label{eqn : 2D good}
\left\{\begin{array}{rcl}
\dot{p}_1^{g} &=& -(\nu_1+b) p_1^g +\lambda_1 p_2^g + \mu_1 + b (p_1^g)^2 \\
\dot{p}_2^{g} &=& -(\mu_2+b) p_2^g +\mu_2 + bp_1^gp_2^g,
\end{array}
\right.
\end{equation}
with initial condition $p_1^g(0) = 0$ and $p_2^g(0)=1$. Note that these boundary values are the only differences between the previous system \eqref{eqn : 2D good}, satisfied by the probabilities $(p_1^g, p_2^g)$, and the system \eqref{eqn : 2D extinction} satisfied by the extinction probabilities $(q_1,q_2)$.
\\

We now turn to the constant rate model. In this case, we can provide a closed formula for the coalescent distribution. Recall the polynomial $Q$ from the previous section
and its two distinct real roots $\alpha < \beta$. 

\begin{prop}
When $\mu_2(\nu_1 - b) -b\lambda_1\not=0$, we have $\beta \not=0$ and then
\begin{equation}
\label{eqn : good Markov case}
W^g(y) = \frac{\mu_2(\nu_1-b)}{\alpha\beta}+ \frac{b\lambda_1\left(\alpha e^{\beta y} - \beta e^{\alpha y}\right)}{\alpha\beta(\beta-\alpha)}\qquad y\ge 0.
\end{equation}
When $\mu_2(\nu_1 - b) -b\lambda_1=0$, we have $\beta=0$, $\alpha = b-\nu_1-\mu_2$, and
\begin{equation}
\label{eqn : good Markov case 2}
W^g(y) = 1+\frac{b\lambda_1}{\alpha^2}\left(e^{\alpha y}- 1 -\alpha y\right)\qquad y\ge 0.
\end{equation}
\end{prop}

\paragraph{Proof.} We first use the method of the homogeneous model. 
In full generality, we can always define the function $F$ as the Laplace transform of the non-negative function $W^g$, that is,
$$
F(s) :=\intgen dy\,e^{-sy} W^g(y) = \intgen dy\,e^{-sy}\left[ W(y) - b\int_0^y dx\, W(y-x)\,P(U>x)\right].
$$
By \eqref{eqn:Laplace} and an integration by parts, we get
\begin{eqnarray*}
F(s) &= & \frac{1}{\psi(s)} - \frac{b}{\psi(s)}\intgen dy\,e^{-sy}\,P(U>y)\\
	&=& \frac{1}{\psi(s)} - \frac{b}{\psi(s)}\left[\frac{1}{s}- \frac{1}{s}\,\intgen dy\,e^{-sy}\,P(U\in dy) \right]\\
	&=& \frac{1}{s\psi(s)}\left[ s-b+b\, E(e^{-sU}) \right]\\
	&=& \frac{1}{s}\ \frac{s-b+b\, E(e^{-sU})}{s-b+b\, E(e^{-s(U+V)})} .
	\end{eqnarray*} 
If we are able to invert this Laplace transform, we get a closed form for $W^g$, and hence for the tail distribution of $H^g$. We have already computed the Laplace transform of $U+V$ in \eqref{eqn:Laplace lifetime}, and trivially $E(e^{-sU}) = \nu_1/(\nu_1+s)$. Plugging these formulae into the general expression for the function $F$ yields 
$$
F(s) = \frac{(s+\nu_1 - b)(s+\mu_2)}{sQ(s)} = \frac{c_0}{s} +  \frac{c_1}{s-\alpha} + \frac{c_2}{s-\beta},
$$
as soon as $\beta\not=0$. Elementary calculus yields
$$
c_0 = \frac{\mu_2(\nu_1-b)}{\alpha \beta},\quad 
c_1 = \frac{(\alpha+\nu_1-b)(\alpha+\mu_2)}{\alpha(\alpha - \beta)}=\frac{b\lambda_1}{\alpha(\alpha - \beta)}, \quad 
c_2 = \frac{(\beta+\nu_1-b)(\beta+\mu_2)}{\beta(\beta-\alpha)}=\frac{b\lambda_1}{\beta(\beta-\alpha)}.
$$
This allows us to invert the Laplace transform of $W^g$ as done in the previous section to get $W^g(y) = c_0 + c_1e^{\alpha y} + c_2 e^{\beta y}$,  which is the announced expression \eqref{eqn : good Markov case}. Similar calculations can be done in the case when $\beta=0$ to get \eqref{eqn : good Markov case 2}. Note that we could have as well used the expression of $W$ computed in \eqref{eqn : all Markov case} and \eqref{eqn : all Markov case 2}, and plugged them into \eqref{eqn : good cpp 2}.\\

Similarly as for the reconstructed tree of all extant species, we can also apply the method used in the Markov model. Indeed, because
$$
W^g(y) =\exp\left(b\,\int_0^y dt\, (1-p_1^g(t))\right)
$$
then 
$$
p_1^g = 1- \frac{W^{g\prime}}{bW^g},
$$
and it is easily seen that $W^{g\prime}$ solves the same second-order, linear differential equation as $W$. The solving details are omitted. \hfill$\Box$

\subsection{The reconstructed tree of representative species}

We now deal with the case of representative species.
Recall from Definition \ref{dfn:order} the ultimogeniture order defined on the set of species born before time $T$, where species $a$ is smaller than species $b$ if their only respective ancestor species $a'$ and $b'$ which were sisters verify that $a'$ is younger than $b'$. Also recall from Definition \ref{dfn:represent} that a representative species is either a good extant species or an incipient extant species which is the first extant descendant of some extinct good species. We want to show that we can again use the contour technique to characterize the reconstructed tree of representative species. Specifically, to ensure that the reconstructed tree of representative species is a coalescent point process, we have to prove that the event that an extant species is representative only depends on the excursion of the contour process that precedes its visit. 

Let $u$ denote some extant species. In the previous subsection, we have argued that the event that an extant species is good or incipient depends only on the corresponding excursion of the contour process. Roughly speaking, an extant species is good iff the last jump of the corresponding excursion has a big enough undershoot $A$, that is, has $U\le A$. Now we claim that $u$ is representative if there is at least one of its ancestor species \emph{first visited during the corresponding excursion}, say $a$, which is good. Indeed, if this last event occurs, then species $u$ is representative by definition, because it is the first extant descendant of all species $a$ satisfying this property, and so is representative of the most recent one among them. Conversely, if $u$ is representative, then its most recent good ancestor species, say $a$ (the species it represents), must be visited for the first time during the excursion. If this was not the case, then $a$ would have another extant descending species previously visited by the contour process, and by definition of the contour process, this species would be smaller than $u$. Then $u$ would not be the smallest extant descending species of $a$, which contradicts the fact that $u$ represents $a$.\\

Now we call $\sigma(u)$ the mother species of $u$, $\sigma^2(u)$ its grandmother species, and so on. We also set $J(u)$ the maximum integer $k$ such that $\sigma^k(u)$ was visited for the first time during the corresponding excursion. We call $A_0$ the age of $u$ at time $T$, and for $i\ge 1$, we call $A_i$ the age at which $\sigma^i(u)$ gave birth to $\sigma^{i-1}(u)$ and $U_i$ the  age at which it turns good. Then $u$ is a representative species iff there is $0\le j\le J(u)$ such that $U_i\le A_i$.
In terms of the contour process, one can detect if an extant species is a representative species if at least one jump of the future infimum of the corresponding excursion has a big enough undershoot, that is, has $U\le A$. We can express this in the following statement, which is the exact analogue of Proposition \ref{prop : good cpp}. 
\begin{prop}
Conditional on at least one representative species extant at time $T$, the reconstructed tree spanned by extant representative species is a coalescent point process. Its associated coalescent distribution is characterized by
$$P(H^r <y) = \frac{P(H<y,\, \exists\, 0\le i\le J(u), U_i\le A_i)}{1- P(H<y, \,\forall\, 0\le i\le J(u), U_i\le A_i)}.
$$
\end{prop}

Unfortunately, we were not able to make a further characterization of the coalescent distribution in the homogeneous model, as was done in Proposition \ref{prop : good cpp 2} for good species, so we now turn to the Markov model.\\

Applying the same arguments as in the last two sections we see that the coalescence time between two consecutive representative species is greater than $y$ if and only if all subtrees that have branched off the lineage of the first one at absolute times belonging in $(T-y, T)$, \emph{do not have any good descending species that has extant descending species by time $T$}. Therefore 
\begin{equation}
\label{eqn : Benelux Markov case}
P(H^r > y) = \exp\left(-\int_0^y dt\,b(t)\, (1-p_1^r(t))\right)\qquad y\ge 0,
\end{equation}
where $p_1^r(t)$ is the probability that a species in the  incipient  stage at absolute time $T-t$  \emph{does not have any good descending species that has extant descending species} at absolute time $T$.

Again, the problem moves to characterizing the function $p_1^r$ and is solved by observing that $p_1^r$ is solution to the following Kolmogorov differential equation, with initial condition $p_1^r(0) =1$,
\begin{equation}
\label{eqn : Benelux ODE}
\dot{p}_1^{r} = -(\nu_1+b) p_1^r +\lambda_1 q_2 + \mu_1 + b (p_1^r)^2,
\end{equation}
where we remind the reader that $q_2(t)$ is the probability that a species in the good stage at absolute time $T-t$ has no descending species at absolute time $T$. Recall that in the constant rate model, this extinction probability is given by Corollary \ref{cor:extinction}. Otherwise, it can be computed thanks to the two differential equations \eqref{eqn : 2D extinction}, so that the coalescent distribution stems from solving a set of 3 differential equations. Recall that the method proposed in Section 3 formally required solving an infinite number of coupled ODEs and does not allow differences in extinction rates between good and incipient species.


\section{Extensions}

\subsection{More stages of incipientness}

Here, we want to extend the Markov model to a model where species can have a fixed number, say $I-1$, of stages of incipientness, before turning good. Namely, we assume that newborn species start in state $1$ as the first stage of incipientness, $2$ the second stage, until they go through stage $I-1$, and finally stage $I$, which is the `good' stage. Assume again that regardless of species status, species give birth (speciate) at the same rate $b$. More specifically, $\lambda_j$ is the rate at which a species of type $j$ becomes type $j+1$ ($1\le j < I$) and $\mu_j$ is the extinction rate of species of type $j$ ($1\le j \le I$). The notation is chosen to be consistent with the constant rate model, which can be obtained by taking $I=2$. Actually, we will now see that this model is a particular case of the homogeneous model treated throughout the paper.\\

Indeed, this model can also be expressed in terms of the durations $V_1,\ldots, V_I$ of successive stages. Start with independent random variables $U_1, \ldots, U_I$, where $U_j$ is an exponential random variable with parameter $\nu_j:=\lambda_j+\mu_j$ if $1\le j\le I-1$, and $\nu_j = \mu_I$ if $j=I$. Also let $\varepsilon_1,\ldots, \varepsilon_{I-1}$ be independent random variables, where $\varepsilon_j$ is a Bernoulli random variable with success probability $\lambda_j/\nu_j$. Set 
$$
N:=\min\{1\le j\le I-1 : \varepsilon_j = 0\},
$$
which is set to $I$ if this last set is empty. Then we can define $V_j:=U_j$ if $j\le N$ and $V_j:=0$ otherwise, as the stage durations of a typical species. More specifically, the species terminates its lifetime in state $N$, its total lifetime duration is $U_1+\cdots +U_N$, and it is in stage $j$ at age $t$ if $V_1+\cdots +V_{j-1} \le t < V_1+\cdots +V_{j}$, $1\le j\le N$. In particular, the species turns good iff $N=I$. Actually, this model is a particular case of the homogeneous model, if we set
$$
U:=V_1+\cdots +V_{I-1}\quad \mbox{ and }\quad V= V_I,
$$
so that we can apply results specifically pertaining to the homogeneous model (equation \eqref{eqn:Laplace} and Proposition \ref{prop : good cpp 2}) to this new model. This involves inverting the Laplace transform \eqref{eqn:Laplace} and compute the distribution function of $U$. We leave the details to the interested reader. Because we do not have specific results in the homogeneous model for the case of representative species, we now explain how to adapt the arguments expanded in the case $I=2$ to the case $I> 2$. Similar reasoning leads to an alternative route as that proposed previously  for the treatment of reconstructed trees spanned by all extant species or by good extant species. This route is detailed explicitly in the next subsection, in the more general setting where some extant species can be missing.\\

Assume again that an extant species is representative iff it is the first extant descendant of some good species. Then equation \eqref{eqn : Benelux Markov case} still holds, namely 
\begin{equation}
\label{eqn : Benelux Markov case I}
P(H^r > y) = \exp\left(-\int_0^y dt\,b(t)\, (1-p_1^r(t))\right)\qquad y\ge 0,
\end{equation}
where $p_1^r(t)$ is the probability that a species in stage $1$ at absolute time $T-t$ has \emph{no good descending species that have extant descending species} at absolute time $T$. The problem is now to characterize the function $p_1^r$. Similarly as in the previous section, the functions $p_j^r$ satisfy the following differential equations, where $p_j^r(t)$ is the probability that a species in stage $j$ at absolute time $T-t$ has \emph{no good descending species that have extant descending species} at absolute time $T$. For any  $1\le j \le I-1$, 
\begin{equation}
\label{eqn : represent  I}
\dot{p}_j^r = -(\nu_j+b) p_j^r +\lambda_j p_{j+1}^r + \mu_j + b p_1^r p_j^r,
\end{equation}
with initial condition $p_j^r(0) =1$, and where  $p_I^r=q_I$ is the probability that a species in the good stage at absolute time $T-t$ has no extant descending species at absolute time $T$.

To compute this extinction probability $q_I$, we let $q_j(t)$ be the probability that a species in stage $j$ at absolute time $T-t$ has no descending species at absolute time $T$, so that the functions $q_j$ satisfy the following differential equations. For all $1\le j \le I-1$,
\begin{equation}
\label{eqn:lastbutnotleast1}
\dot{q}_j = -(\nu_j+b) q_j +\lambda_j q_{j+1} + \mu_j + b q_1q_j,
\end{equation}
and for $j=I$,
\begin{equation}
\label{eqn:lastbutnotleast2}
\dot{q}_I= -(\mu_I+b) q_I +\mu_I + bq_1q_I,
\end{equation}
with initial conditions $q_j(0) = 0$ for all $1\le j\le I$, which are the analogues to equations \eqref{eqn : 2D extinction} satisfied by extinction probabilities in the case $I=2$.
 To sum up, there are $2I-1$ differential equations to solve in order to get the coalescent distribution \eqref{eqn : Benelux Markov case I}. First, one has to solve the previous system of $I$ differential equations to compute the extinction probabilities $q_j$, and then plug $q_I=p_I^r$ into the system \eqref{eqn : represent  I} of $I-1$ differential equations to get $p_1^r$.

\subsection{Missing species}
\label{subsec:missing}

In this subsection, we complete the calculations made in Subsection \ref{subsec:LF missing} for trees with missing species under a binomial sampling scheme, when sampling probability depends on species status.
We will treat this case only in the Markov model, but for the sake of completeness, we will consider the full generality of $I$ stages of incipientness, as in the previous subsection. 

From now on, we define $\rho_i$ as the probability of being sampled at $T$ for an extant species in stage $i$. We wish to compute the likelihood of the reconstructed tree of all \emph{sampled} species or of \emph{sampled} representative species. Notice that the reconstructed tree of (sampled or not) good species can be seen as the reconstructed tree of all sampled species in the special case when $\rho_i=0$ as soon as $i\not=I$ (recall that $I$ is the stage of good species). 

The reconstructed tree of all sampled species (resp. of all representative species) is again a coalescent point process, and the common density $f$ (resp. $f_r$) of its typical node depth $H$ (resp. $H_r$) satisfies the same ordinary differential equations as previously, but with  different initial conditions. Let us give a conclusive, self-contained  summary of these results. 

We first modify slightly the definitions of the quantities $q_j(t)$ and $p_j^r(t)$. We now let $q_j(t)$ stand for the probability that a species in stage $j$ at absolute time $T-t$ has no descending species \emph{sampled} at absolute time $T$. The functions $q_j$ still satisfy the differential equations \eqref{eqn:lastbutnotleast1} and \eqref{eqn:lastbutnotleast2}, but with initial conditions $q_j(0) = 1-\rho_i$ for all $1\le j\le I$. Recall that it is possible, for example, to recover the probability $p_j^g$ that a species in stage $j$ at absolute time $T-t$ has no \emph{good} descending species at absolute time $T$, by taking $\rho_i=1$ if $i\not=I$ and $\rho_I=0$.

Now we let $p_j^r(t)$ be the probability that a species in stage $j$ at absolute time $T-t$ has no good descending species that have extant descending species \emph{sampled} at absolute time $T$. Then the functions $p_j^r$ still satisfy the differential equations \eqref{eqn : represent  I}, again with $p_I^r=q_I$, and with the same initial conditions $p_j^r(0) =1$.

Now similarly as in Section \ref{sec:all}, we set $g(y):=P(H>y)$ and $g_r(y):=P(H_r>y)$, so that $f=-\dot{g}$ and $f_r=-\dot{g}_r$. It is easy to see that
$$
P(H > y) = \exp\left(-\int_0^y dt\,b(t)\, (1-q_1(t))\right)\qquad y\ge 0,
$$
and that 
$$
P(H_r > y) = \exp\left(-\int_0^y dt\,b(t)\, (1-p_1^r(t))\right)\qquad y\ge 0,
$$
so that 
$$
\dot{g} = -b(1-q_1) g\quad\mbox{ and }\quad \dot{g_r} = -b(1-p_1^r) g.
$$
Each of the previous two equations can be solved simultaneously with those satisfied by the probabilities $(q_i)$ and/or $(p_j)$. Then use $f=-\dot{g} = b(1-q_1) g$ and $f_r=-\dot{g}_r=b(1-p_1^r) g$.

\paragraph{Acknowledgments.} AL was financially supported by grant MANEGE `Mod\`eles Al\'eatoires en \'Ecologie, G\'en\'etique et \'Evolution' 09-BLAN-0215 of ANR (French national research agency). HM was funded by the CNRS and ANR grant ECOEVOBIO-CHEX2011. RSE was financially supported by the Netherlands Organisation for Scientific Research (NWO) through a VIDI grant.

\bibliographystyle{apalike}
\bibliography{myref} 



\end{document}